\documentclass[a4paper,10pt]{article}
\usepackage{amsmath} % Espressioni matematiche
\usepackage{graphicx} % Inserimento immagini e grafiche varie
\usepackage{subfig}
\usepackage{float} % Posizionamento immagini
\usepackage{xcolor}
\usepackage[utf8x]{inputenc} % Riconoscimento lettere accentate
\usepackage{lmodern}
\usepackage{amsmath}
\usepackage{amsfonts}
\usepackage{amssymb}
\usepackage{chemarrow}
\usepackage{tabularx,ragged2e,booktabs,caption}
\usepackage{bm}
\usepackage{etoolbox}
\usepackage{calc}% http://ctan.org/pkg/calc
\usepackage[american]{babel}
\usepackage[sort,round]{natbib}
\usepackage{graphicx}
\usepackage{cite}
\usepackage{hyperref}
\usepackage{tikz}
\usepackage{accents}
\usepackage{titlesec, blindtext, color}
\usepackage[toc,page]{appendix}
\usepackage{authblk}

\begin{document}
% mie definizioni
\newcommand{\vi}{``}
\newcommand{\virgolet}[1]{``#1''}
 
\newcommand{\ubar}[1]{\underaccent{\bar}{#1}}

%delimitatori vari
\newcommand{\del}[3]{\left #1 #2 \right #3}
 \newcommand{\bra}[1]{\del \langle #1  | }
 \renewcommand{\k}[1]{\del| {#1} \rangle }
 
%colori
\xdefinecolor{bordo}{rgb}{0.5,0,0}
\xdefinecolor{blu}{rgb}{0,0,0.4}
\xdefinecolor{giallo}{rgb}{1,1,0.5}
\xdefinecolor{gial}{rgb}{1,1,0.2}
\xdefinecolor{lavendar}{rgb}{0.8,0.6,1}
\xdefinecolor{olive}{cmyk}{0.64,0,0.95,0.2}
\newcommand{\cfbox}[2]{%
    \colorlet{currentcolor}{.}%
    {\color{#1}%
    \fbox{\color{currentcolor}#2}}}
%somme
\newcommand*\circled[1]{\tikz[baseline=(char.base)]{
            \node[shape=circle,draw=orange,inner sep=2pt] (char) {#1};}}
    
\newcommand{\angstrom}{\mbox{\normalfont\AA}}
\newcommand{\abs}[1]{\lvert#1\rvert}
\newcommand{\blu}{\frac{\sum\limits_{s \in S} s^2}{\sum\limits_{p \in P} p^2}}
\newcommand{\sumns}{\sum\limits_{s=1}^{N_{\text{sect}}}}
\newcommand{\sumj}{\sum\limits_{j= \; b_{s-1}+1}^{b_s}}
\newcommand{\sumjl}{\sum\limits_{j=1}^s}
\newcommand{\sumjb}{\sum\limits_{j=1}^{b_1}}    
\newcommand{\sumin}{\sum\limits_{i=1}^n}
\newcommand{\sumc}{\sum\limits_{c}}
\newcommand{\sump}{\sum\limits_{p}}
\newcommand{\sumcnc}{\sum\limits_{c}^{N_c}}
\newcommand{\sumpnp}{\sum\limits_{p}^{N_p}}
\newcommand{\sumrpr}{\sum\limits_{r'}}
\newcommand{\sumppr}{\sum\limits_{p'}}
\newcommand{\sumsp}{\sum\limits_{s'}}
\newcommand{\sumpcpr}{\sum\limits_{c'p'}}
\newcommand{\sumscpr}{\sum\limits_{'p'}}
\newcommand{\ig}{\int_{0}^1}
\newcommand{\sums}{\sum\limits_{s}^{N_{\text{sect}}}}

\title{Economic Development and Inequality: a complex system analysis \\ \Large -- Working Paper --}
% \author{Angelica Sbardella, Emanuele Pugliese, and Luciano Pietronero}
% \affil{ISC-CNR, Via dei Taurini 19, 00185, Rome, Italy}
\author[1]{Angelica Sbardella}
\author[1]{Emanuele Pugliese} 
\author[2]{Luciano Pietronero}

\affil[1]{ISC-CNR - Institute of Complex Systems, Rome, Italy}
\affil[2]{Department of Physics, Sapienza Universit\`{a} di Roma, Rome, Italy}
% % 
% % Dipartimento di Fisica, Sapienza Universita di Roma, P.le Aldo Moro 2, 00185, Rome, Italy
\date{May, 2016}
\maketitle
% %\listoffigures
\begin{abstract}
By borrowing methods from complex system analysis,  in this paper we analyze the features of the complex relationship that links the development and the industrialization of a country to economic inequality. In order to do this, we identify industrialization as a combination of a monetary index, the GDP per capita,  and a recently introduced measure of the complexity of an economy, the Fitness.  At first we explore these relations on a global  scale over the time period 1990--2008 focusing on two different dimensions of inequality: the capital share of income and a Theil measure of wage inequality. In both cases, the movement of inequality  follows a pattern similar to the one theorized by Kuznets in the fifties. We then narrow down the object of study ad we concentrate on wage inequality within the United States. By employing data on wages and employment on the approximately 3100 US counties for the time interval 1990--2014, we generalize the Fitness-Complexity algorithm for counties and NAICS sectors, and we investigate wage inequality between industrial sectors within counties. At this scale, in the early nineties we recover a behavior similar to the global one. While, in more recent years, we uncover a trend reversal: wage inequality monotonically increases as industrialization levels grow. Hence at a county level, at net of the social and institutional factors that differ among countries,  we not only observe an upturn in inequality but also a change in the structure of the relation between wage inequality and development.

\end{abstract}

\section{Introduction}
%EMANUELE: RIVEDIAMO INTRO E AGGIUNGERE LE CITAZIONI MANCANTI  E METTERE RIF A HIDALGO
The inequality of income has been a central issue in economic and political debates since the beginning of the economic as a discipline \citep{smith,marx}. 
Our aim is to look at how the industrialization of a country affects the income inequality of its population. \footnote{Even if we will not emphasize this aspect, it is also worth mentioning how the opposite causal relationship, the role of inequality in explaining the development of a country, has been theorized. Acemoglu in \citep{acemoglu12,acemoglu13} relates increased political and economic inclusivity in a society with increased innovation and how Roseinstein-Rodan \citep{rosenstein43} and Murphy, Shleifer and Vishny \citep{murphy88} \vi Big Push" kinds of models predict a role of a sizeable middle class in producing enough demand for industrialization.}
For this purpose, first we need to quantitatively define  the  income inequality and the industrialization of a country. It is possible to define inequality in several ways. Not only different measures of the dispersion of a distribution are possible --the Gini coefficient, the Coefficient of Variation, the Theil index, the Herfindel Index, the ratio between the top $10\%$ and the bottom $10\%$ and so on -- but also the object of inequality can vary.  Total income can be naturally divided into two parts, depending on the source of income. These are labor income and capital income, respectively wages or capital gains.
Consequently, economic theory about income inequality can be broadly divided into two main streams according to the different objectives of analysis: i) the ratio between capital or labor income and total income -- the capital or labor share -- present in a literature stemming from \citealp{marx} to \citealp{piketty-capital}; ii) the inequality in the distribution of total, capital or labor income  (\citealp{pareto, gini, katz-goldin09, kuznets55,galbraith01}, etcetera). %BISOGNO DI REF
In this paper we will focus on the inequality of labor income, but we will show that our new way to define industrialization allows us also to say something  about the evolution of capital share in industrializing economies.

In mainstream economics the theoretical model to explain both the capital share of income and the inequality of labor income is based on the idea that a worker's wage is completely determined by her skill level, the relative abundance of capital and labor, and by the technological landscape.  
The basic principles of supply and demand rule the labor market, just like in any other market. The consequence of this view is that the increasing pay inequality observed in the last three decades would be  fundamentally due to the so called \vi skill--biased technical change''\citep{katz-goldin09}. 

However, many arguments relate macroeconomic development to inequality through the relationships between development and education (skills), as well as demographic and investment variables.
A brief review of the main arguments and predictions is due at this point.

Piketty \citep{piketty-capital} found that the capital share of income is growing in the developed world and theorizes that it will further increase. Indeed, he connects the capital share to the growth of developed country's economy, predicting an inevitable increase in the capital share, while the growth of  developed countries population and economy declines.  Therefore, Piketty's \vi Laws of Capitalism" predict that, with the end of the industrialization process, economic growth would naturally decline and inequality would inevitably rise.

A more optimistic argument is made in the seminal work of Kuznets \citep{kuznets53,kuznets55}.  The pioneering work of Kuznets was developed in 1955 and was based on data concerning only the US between 1913 and 1948.  
According to Kuznets, in the process of economic development --in particular in the inter-sectoral transition from an agriculture based economy to an industrial one--  a general relation which ties wealth and inequality to pay subsists. 
When the passage towards industrialization takes hold there is an injection of cheap rural labor to the cities that holds down wages. At the same time, whoever owns capital has more investment possibilities and increasing the return on investments. These factors increase both the capital share of income and the inequality of labor income as the economy starts its industrialization. This provokes a wide urban-rural gap and a consequent rise of inequality. Kuznet's innovation stands in the relationship between economics and policies. 
He predicts that, when capitalism enters in its advanced phase, the labor force bargains to improve pay and work conditions through social struggle, ultimately leading to a strengthening of the welfare state, and a process of democratization is triggered in which modern industrial relations are created and inequality decreases.
%All elements that pave the way for social democracies in which, almost in a mechanic way, pay inequality is supposed to decrease. 
The overall increase and decrease of inequality in the development of a country defines the inverted-U \vi Kuznet's curve''.

%The pioneering work of Kuznets was developed in 1955, a far optimistic period in which the non-communist world considered growth the cure for all economic diseases, and it was based on data concerning only the US between 1913 and 1948.
Galbraith  looks at the inequality in labor income through a Theil index for more recent country time series, and failing to recover the whole Kuznets' curve, finds only its leftmost part \citep{galbraith01,galbraith07}.

The different theoretical predictions are difficult to compare with empirical results since many terms are broadly defined.  It is particularly difficult to identify industrialization, both its first phase -- the moment of increased opportunities for workers to educate themselves and capitalists to invest -- and its second phase  -- the following moment predicted by Kuznets in which people's basic necessities are satisfied and government policy starts including inequality as an issue.
In \citep{pugliese15} the authors define the metric for this evolution as a combination of a monetary value, GDP per capita, and a non monetary value, Fitness, the complexity of a  country's economy. The measure of this apparently intangible feature has been tackled in \citep{hidalgo09,tacchella12}, analyzing at the export network of countries.

In this paper, we will firstly generalize the metric developed by Tacchella et al. for different networks and then we will use an approach similar to the one of Pugliese et al. to define the industrialization of the country in the GDP per capita-Fitness phase space.  %the emphasis here is on the patterns
We will finally be able to look at how inequality maps in this space.
Furthermore, we will analyze the scale properties of the inequality process.
In fact, we will study the behaviors of the inequality process among countries within and within a single country, the United States. Within the US we will measure labor inequality between industrial sectors within counties by using a Theil index, an entropy-based measure naturally decomposable among population groups, which will allow us to study the system and its component in a consistent way. 
%Using a Theil index to measure labor inequality, an index that is naturally decomposable, therefore allowing us to study a system and its parts in a consistent way, we will look to see if the inequality process behaves in the same way among countries and among their parts, focusing in this case on the counties of the United States.
To do so,  we will employ data on wages and employment levels of the US at a county level; we will compute county Fitness by looking at the localization of industrial sector in each county, as it is common in geographical economics \citep{kim95}.
We will show how the trends of the distribution of wages change drastically at different scales and how those trend vary over time.

The remainder of this paper is structured into four sections.  Section 2 describes the sources of data that we use, both on a global scale and on a regional scale, within the United States. Section 3 introduces our variable of interest, Fitness as a measure of economic complexity and a Theil measure for the inequality between industrial sectors. Section 4 shows the results of our study. And, finally, Section 5 presents our conclusions.

\section{Sources of data}\label{sec:data}
We carry out an analysis on the possible correlation between labor inequality, Fitness and the level of GDP per capita or labor income grouped by economic sectors. In a first phase, we will focus on a pooled analysis of a panel of different countries. While, in a second phase, we will narrow the field of our investigation and we will examine cross-sections of the US counties.
In the followings we briefly illustrate the employed data.

\subsection{Penn World Table 8.1 (PWT)}
\textit{Penn World Table 8.1} (PWT) is a database produced by the \textit{University of Groningen} and the \textit{University of Pennsylvania} and it provides levels of income, output, input and productivity, covering 167 countries for the period 1950-2011 \citep{pwt}. In particular, we are interested in data on Expenditure-Side Real GDP at current purchasing power parities, populations and on estimates of the share of labor compensation in GDP.

\subsection{Export Fitness}
As a measure of Economic Complexity we take the Export Fitness of countries, a dimension recently proposed in \citep{tacchella12}. For a more detailed discussion of Fitness we refer to the Section \ref{fitness-section}.
The Export Fitness data-set covers a number of countries varying slightly between 145 and 148, over the period 1995--2010. The considered export volumes comprise 1131 products and are taken from the BACI database \citep{baci}. Exported commodities are classified according to the four-digit  Standard International Trade Classification (SITC), Revision 2.

\subsection{University of Texas Inequality Project (UTIP-UNIDO)}
As a measure of pay inequality worldwide we employ UTIP-UNIDO, a global data-set produced by the \textit{University of Texas Inequality Project} with the support of INET \citep{utip-data}. It encompasses a Theil measure of pay inequality across manufacturing sectors covering 167 countries during the period 1963--2008.  
The data on wages is drawn from the Industrial Statistics database published annually by the United Nation Industrial Development Organization (UNIDO), where industrial sectors are categorized according to the International Standard Industrial Classification (ISIC) at a two or three-digit aggregation level.
% It has a total of 4054 observations 

\subsection{Bureau of Labor Statistics (BLS)}
The data on employment and wages regarding the United States is taken from the Quarterly Census of Employment and Wages (QCEW) data-set of the US Bureau of Labor Statistics for the period 1990--2014 \citep{bls}. %, the U.S. Department of Labor and the state Employment Security Agency.
In the QCEW industries are labeled by the North American Industry Classification System (NAICS), a standard method used in Canada, Mexico, and the United States to classify business establishments according to types of economic activity. The NAICS numbering system employs a six-digit code at the most detailed industry level. %The first two digits specify the largest business sector, the third digit describes the sub-sector, the fourth digit the NAICS industry group, the fifth digit the NAICS industries and the sixth digit the national industries.%across the U.S.. 
For the approximately 3100 American counties, the QCEW provides employment and earnings information at a county-level by six-digit NAICS industry and by ownership sector.
This data is also aggregated to annual levels, to higher industry levels %(NAICS industry groups, sectors and supersectors) 
 and to higher geographical levels (national, states).  In our analysis we look at the private sector and three-digit NAICS industries on a county-level. 

\section{Variables of interest}\label{variables}

\subsection{Fitness as a measure of Economic Complexity}\label{fitness-section}
Economists have long held the view that only one statistic, the measure of GDP, could account for national progress and development. Over recent years, however, monetary indexes such as GDP or its equivalents have been widely criticized and alternative or complementary indicators have been introduced \citep{costanza09,stiglitz-sen09,diener97}. %(Costanza et al., 2009, Stiglitz and Sen, 2009, Diener and Suh, 1997).
In the framework of Economic Complexity there is a vivid debate on how the notion of complexity might integrate the information given by monetary indexes. In a recent strand of literature Tacchella et al. introduced a new dimension to assess the economic complexity of a country, Fitness $F$ \citep{tacchella12, tacchella13, cristelli13, cristelli15}.  $F$ is an indirect measure of the manufacturing capabilities of a country. The capabilities are representative of the underlying social and economic structure of a society, and are the sum of all those national characteristics that enable a country to produce and export goods. 
By describing the international goods market as a bipartite network of countries and products, the measure proposed by Tacchella et. al. provides a ranking of the development  potential of countries quantifying the diversification and the complexity of their export baskets. In designing a method to extract information from the country-product network, the authors drew inspiration from the Google Page Rank algorithm \citep{google-pr}. They created a non-linear coupled map in which a fixed point defines a metric for the Fitness of countries and the Complexity of their products.
 
%As shown in various recent works by Tacchella et al. (\citep{tacchella12}, \citep{tacchella13}) and Cristelli et al. (\citep{cristelli13}, \citep{cristelli15}), Economic Complexity has a role in evaluating  the competitiveness of countries in international trade. But we believe that it could also be a useful tool for examining the economical processes within a country. We would like to generalized the metric obtained for  export; a metric that could provide the Fitness and the Complexity of an entire economy for a generic geographical unit. If, instead of looking at international trade,  we examine the whole productive system of a certain region, we could find a metric that takes into consideration the economic activities that compose all the domestic market of a country and permit export. In fact, since the Fitness--Complexity algorithm defines an intensive metric, it is possible to generalize its definition to regions -where by region we mean a generic geographical unit that can be a sub--division of the same country or a macro--area

Given that this map -- the Fitness-Complexity algorithm -- defines an intensive metric, it is possible to generalize its definition, taking into consideration a bipartite network of regions and economic activities -- in which  by region we mean a generic geographical unit that can be a sub-division of the same country or a macro-area. By doing so, it would be possible to give information on the whole economy of a country, not only on its manufacturing capabilities. 
%A different application of the Fitness-Complexity algorithm has already been tested in \citep{napolitano15} CITAZIONE ARTICOLO BREVETTI for countries and patents. 
In this paper we focus on labor income: we consider the US counties as geographical units and the NAICS sectors as economic activities. Data on wages and employment is taken from the Quarterly Census of Employment and Wages data set of the Bureau of Labor Statistics described in Section \ref{sec:data}. In a work soon to be submitted, we are exploring in depth this generalization of the algorithm for the US and its application on different geographical scales (i.e. states and counties) and different industrial aggregations employing data on NAICS industries from two to six-digits.
%\textcolor{red}{CITARE PROX ARTICOLO E DIRE CHE SI FARA' ANALISI APPROFONDITA DI COME FUNZIONA L'ALGORITMO PER LE CONTEE}
As a reasonable indicator of productive system \vi dimensions'',  we employ the total earnings ${W}_{rs}$ of  the workers of sector $s$ in region $r$ over one year. Thus, if we take $N_s$ economic sectors and $N_r$ regions,  we can build a sector-region matrix $\hat M$ of dimension $ N_r \times N_s $. The matrix element $M_{rs}$ specifies if a sector $ s$ is present or absent in the region $r$, with the entries being the wage volume ${W}_{rs}$ for $r=1,\cdots, N_{r}$ and $s=1,\cdots, N_s$.
Along the same lines of the Fitness-Complexity algorithm method, in order to make the region-sector matrix  $\hat M$ binary, we use the Revealed Comparative Advantage \citep{balassa} of a sector $ s$ in a region $ r$:
\begin{equation}\label{rcars}
RCA_{rs}= \dfrac{{W}_{rs}}{\sumrpr {W}_{r's}}/\dfrac{\sumsp {W}_{rs'}}{\sum\limits_{r's'} {W}_{r's'}} 
\end{equation}
the ratio of the wage share of sector $s$ in region $r$ to the wage share of sector $s$ in the whole geographic area under consideration.  We identify a region $r$  as competitive in a particular industrial sector $s$ if $RCA_{rs} \geq 1$.  
This version of the Revealed Comparative Advantage is closely related to a classical tool of economic geography, the Location Quotient (LQ).  With a similar notation to the one in the definition of the $RCA_{rs}$ in Eq.\ref{rcars}, $LQ_{rs}$ of sector $s$ for region $r$ is a ratio that allows us to compare the distribution of employment by industrial sector in an area wiwithh a reference distribution, which in general is the national one. And if $E_{rs}$ is the number of workers in sector $s$ for region $r$, the Location Quotient is expressed by the formula:
\begin{equation}
LQ_{rs}= \dfrac{{E}_{rs}}{\sumrpr {E}_{r's}}/\dfrac{\sumsp {E}_{rs'}}{\sum\limits_{r's'} {E}_{r's'}}.
\end{equation}
When $LQ_{rs}=1$ the share of employment in sector $s$ is equal for the regional and the national economy; while when $LQ_{rs}<1$ the share of regional employment is less than it is in the national case; and vice versa for $LQ_{rs}>1$. In this light, if $w_{rs}$ is the average wage of $s$ in $r$, Eq.\ref{rcars} can be rewritten in this way:
\begin{equation}\label{rcarslq}
RCA_{rs}= LQ_{rs} \cdot \dfrac{{w}_{rs}}{\sumrpr {w}_{r's}}/\dfrac{\sumsp {w}_{rs'}}{\sum\limits_{r's'} {w}_{r's'}}.
\end{equation}
% We consider country c to be a competitive exporter of product p if its RCA is larger than some threshold value, which we take as 1 as in standard economics literature; previous studies have verified  that small variations around such threshold do not qualitatively change the results %given by the semipositive matrix
In order to remove any trivial correlations with wage volumes, in matrix  $\hat{M}$ we report only whether a sector is present or not in a region.  When $s$ is present  in $r$, we assign to the matrix element $M_{rs}$ the value $1$, and the value $0$ otherwise. Hence, the binary region-sector matrix will have generic elements of this kind:
\begin{eqnarray}\label{mrs}
M_{rs}=
\begin{cases}
1 & \text{ if $\,RCA_{rs} \geq 1$}\\ 
0 & \text{ otherwise.}\\
\end{cases}
\end{eqnarray}

\begin{figure}[H]
\begin{center}
\includegraphics[scale=0.5]{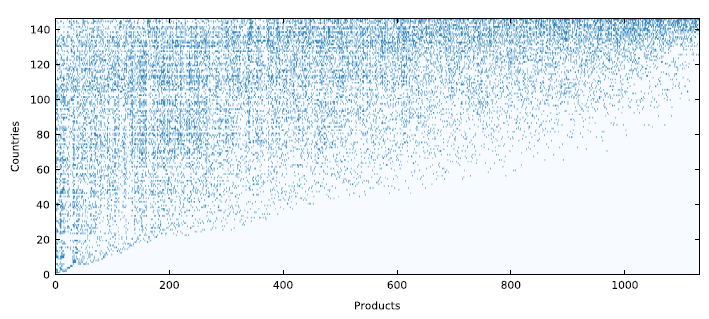}
\end{center}
\caption{\small The binary matrix of countries and products built from the worldwide export flows of BACI  data-set for 1998, with $N_{countries}=147$ and $N_{products}=1131$ \citep{tacchella13}. Products are categorized according to the Harmonized System 2007 at four digit coarse-graining and the adopted digitalization criterion is Balassa's Revealed Comparative Advantage.  By sorting the columns of the matrix by increasing Fitness and the rows by increasing Complexity, the matrix acquires a triangular-like shape.  As it turns out, countries with more diversified export baskets are more competitive, while countries specialized in a few products --which generally are also exported by every other country-- are the poorest.}\label{mcp-pil}
\end{figure}

\begin{figure}%[H]
\centering
\subfloat[]{
 \hspace{-17pt}
  \includegraphics[width=65mm]{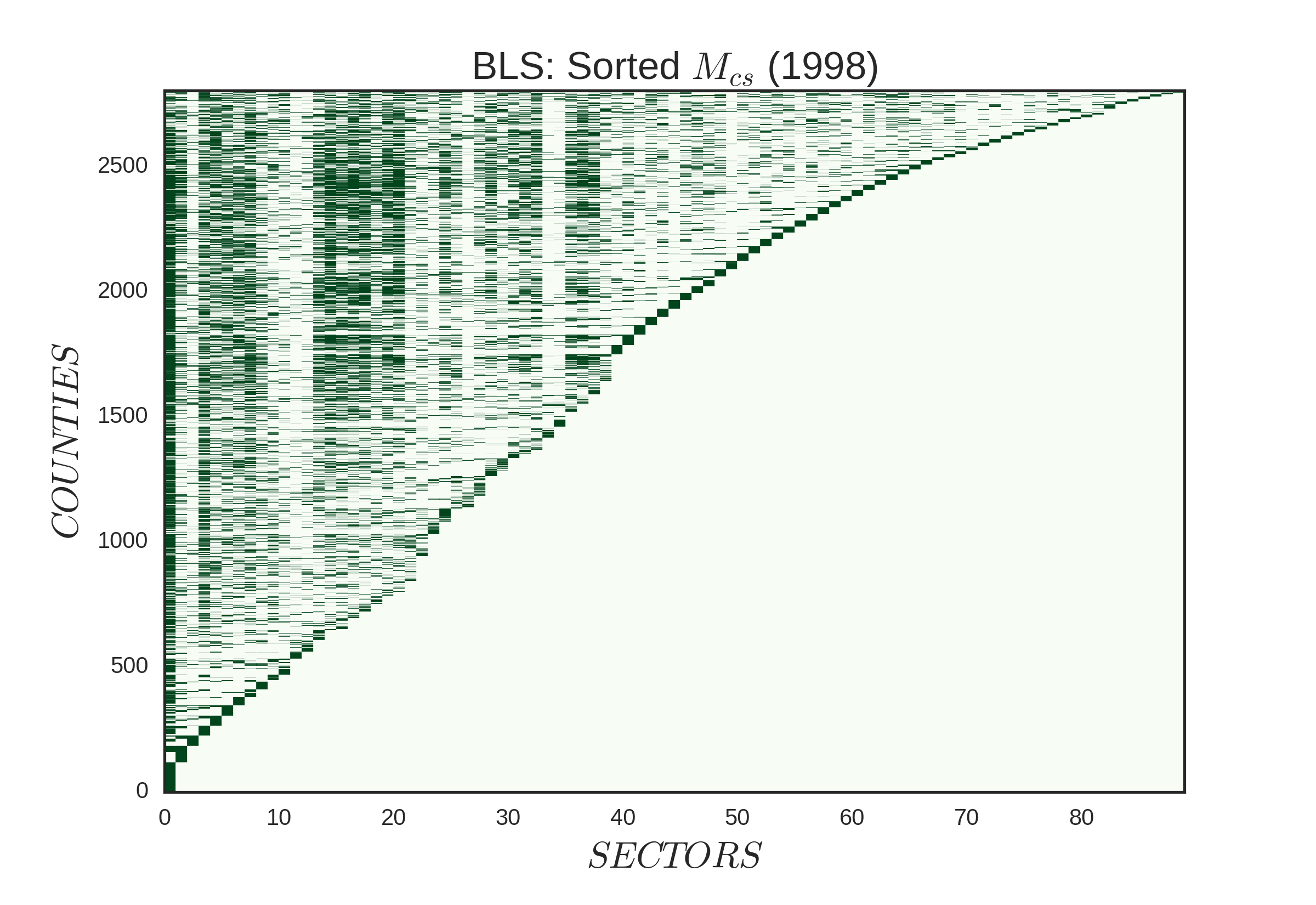}
}
\hspace{0mm}
\subfloat[]{
 \hspace{-17pt}
  \includegraphics[width=65mm]{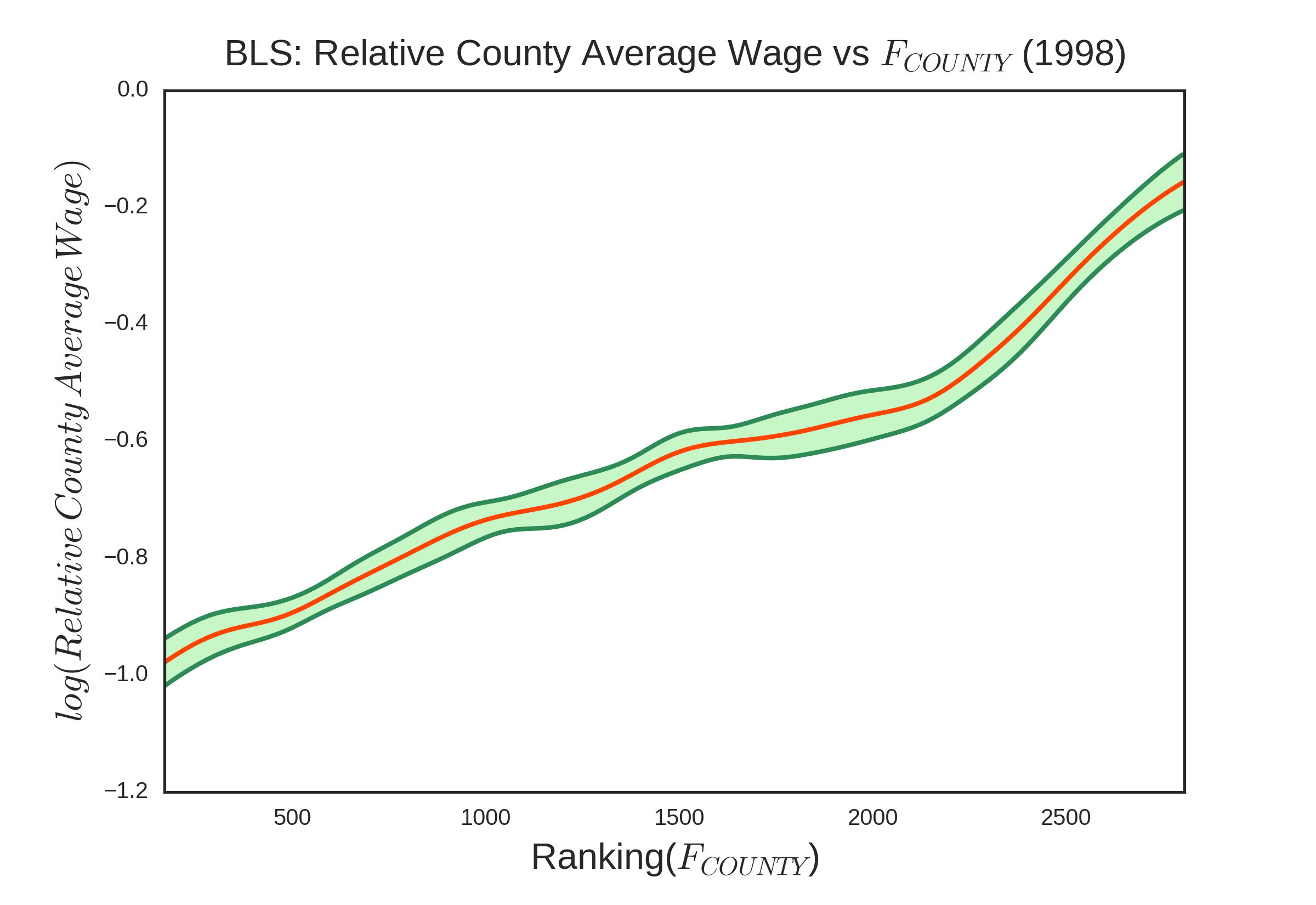}
}
\subfloat[]{
 \hspace{-17pt}
  \includegraphics[width=65mm]{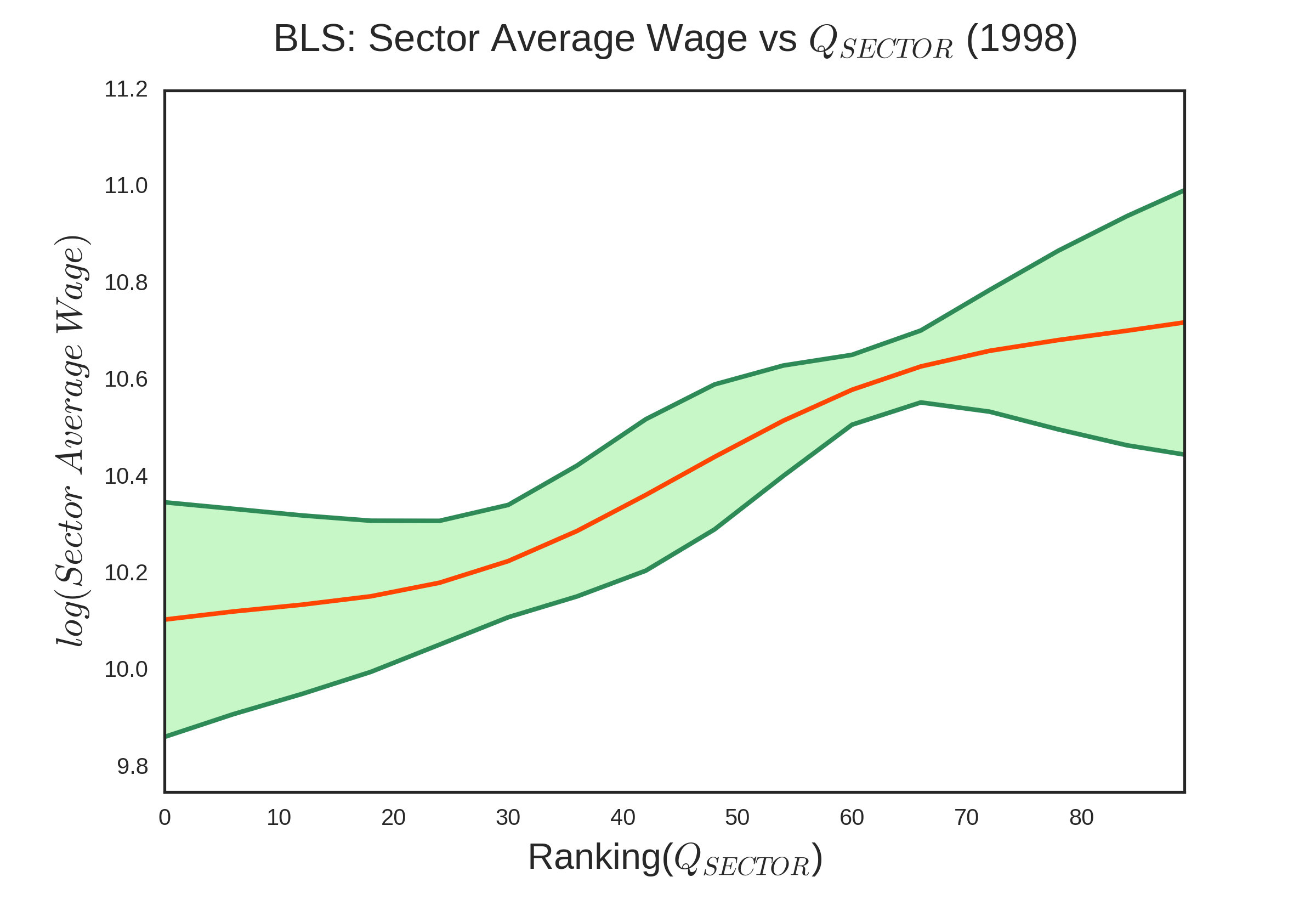}}
\caption{\small The three figures refer to the US counties in 1998 and are realized through considering data on wages and employment from Quarterly Census of Employment and Wages data set of the Bureau of Labor Statistics. In this case, sectors are categorized according to the NAICS industry classification at three digit aggregation level, with $N_{counties}=2805$ and $N_{sectors}=89$.  (a): The county-sector matrix $\hat{M}$ is built from the volume of sectoral wages in US counties. By employing the same ranking principle used for the country-product matrix, also here $\hat M$ acquires a triangular shape. Hence, the path towards diversification in production of goods and services is not only taken, as we have seen in \ref{mcp-pil}, by high Fitness countries but also by the most complex and diversified US counties.(b): The relation between Relative County Average Wage and County Fitness. The red line represents a kernel estimation of County Average Wage versus $F_{COUNTY}$ and the green shadowed area shows a $90\%$ confidence interval of the expected value. On a country level the monetary counterpart of Fitness was the GDP per capita, while here we adopt the Average Wage as a proxy of the wealth of a county. The trend is concordant with the one found comparing countries, in the sense that there is a non-linear relationship between the two variables but in general as $F_{COUNTY}$ grows the Average Wage increases. As in the export case, $F_{COUNTY}$ shows some deviations from the monetary metric that can give us some information on the overall pay distribution. (c): The relationship between national Sector Average Wage and Sector Complexity. As in the previous case, the red line is a kernel estimation of Sector Average Wage versus $Q_{SECTOR}$ and the green area is a $90\%$ confidence interval computed with bootstrap. As it seems reasonable, Sector Average Wage grows with Sector Complexity $Q_{SECTOR}$.}\label{fig:us-mcs-cavw-f-savw-q98}
\end{figure}

We build an empirical $\hat M$ from the volume of wages in the us counties over the interval 1990--2014. In Fig.\ref{fig:us-mcs-cavw-f-savw-q98} section (a) we report the 1998 matrix. Ranking the columns of the binary county-sector matrix by increasing County Fitness and the rows by increasing Sector Complexity, $\hat M$ assumes a triangular-like shape, as it did in the analysis of international export (Fig.\ref{mcp-pil}). Therefore, the most fit counties are the most diversified --almost all sectors are present in their productive system-- while the regions specialized on few sectors are the less competitive. So, diversification seems an essential ingredient also in explaining competitiveness and wealth within the United States.
% Basic Ricardian paradigm would predict an almost block-diagonal matrix. The matrix instead has a triangular-like shape. On one hand the most diversified countries result to be the most competitive while the ones specialized on those few products exported by almost every country are the poorest. On the other hand this implies that specialization is an uncommon strategy among most successful countries pointing out that diversification is a more important element to explain the competitiveness and the wealth of countries. 
Thus, going back to the formal description, by following the scheme of Tacchella et al. from the matrix $\hat M$ it is possible to obtain an intensive metric that measures the region Fitness $F_r$ as the diversification weighted by the Sector Complexity, and the Sector Complexity ${Q_s}$ as the diversification bounded by the Fitness of the less competitive county in which the sector is present. Between these two variables a non-linear coupling holds, which is expressed in the Fitness-Complexity iterative algorithm (defined in Eq.\ref{pilrs1} and Eq.\ref{pilrs2}) in which, at every step, $F_r$ and ${Q_s}$ are evaluated and normalized:

\begin{eqnarray}\label{pilrs1}
\begin{cases}
\widetilde{F}_r^{(n)}=\sum_s M_{rs} Q_s^{(n-1)} & \\ \\
\widetilde{Q}_s^{(n)}=\dfrac{1}{\sum_r M_{rs} \dfrac{1}{F_r^{(n-1)}}} \\
\end{cases}
\begin{cases}
F_r^{(n)}=\dfrac{\widetilde{F}_r^{(n)}}{<\widetilde{F}_r^{(n)}>}  & \\ \\
Q_s^{(n)}=\dfrac{\widetilde{Q}_s^{(n)}}{<\widetilde{Q}_s^{(n)}>}.\\ 
\end{cases}
\end{eqnarray}

With initial conditions:
\begin{eqnarray}\label{pilrs2}
\begin{cases}
\sum_r F_r^{(0)}=1  \hspace{6pt}\forall r   \\ \\
\sum_s Q_s^{(0)}=1   \hspace{6pt}\forall s.\\ 
\end{cases}
\end{eqnarray}

% \begin{figure}[H]
% \begin{center}
% \includegraphics[scale=0.5]{US/USA-Sorted-Mcs-1998-wage.png}
% \end{center}
% \caption{\small (c): The county-sector matrix $\hat{M}$ built from the volume of sectoral wages in U.S. counties for 1998 by employing the Quarterly Census of Employment and Wages data set of the Bureau of Labor Statistics. In this case, sectors are categorized according to the NAICS industry classification at three digit aggregation level, with $N_{counties}=2805$ and $N_{sectors}=89$. By employing the same ranking principle used for the country-product matrix,  also here $\hat M$ acquires a triangular shape. Hence, the path towards diversification in production of goods and services is not only taken, as we have seen in \ref{mcp-pil}, by high Fitness countries but also by the most complex and diversified US counties.}\label{mcsus98}
% \end{figure}

%us-fitness-complexity

The iteration of the coupled equations leads to a fixed point which has been proved to be stable and non dependent on initial conditions \citep{tacchella12}.  The fixed point defines a non-monetary metric which quantifies the Fitness of the economic system in analysis and the Complexity of its economic sectors. This generalized Fitness-Complexity metric takes into account the overall economy, and also those sectors that have a role only in the domestic market, such as the service industry.  In Fig.\ref{fig:us-mcs-cavw-f-savw-q98}, the outcomes for the US counties in 1998 are shown. Section (b) displays County Fitness versus Relative County Average Wage;  while, section (c)  shows Sector Complexity versus Sector Average Wage. The results are concordant with those of Tacchella et al. Therefore, the redefinition of the Fitness-Complexity algorithm presented in this section appears a solid instrument to study the comparative development of US counties.

% the latter used as the monetary counterpart of the former. The relation reproduces quite faithfully the known dependency of GDP per capita from Country Fitness, as widely explained in \citep{cristelli15}.  In fact, as in the export analysis, the Fitness of the American counties shows some deviations from the chosen monetary metric and those deviations can be highly informative on the county competitiveness. For instance, a county which has low or intermediate Average Wage and high Fitness potentially has more chances to grow than a county that has the same wage-level but lower $F_C$. Indeed, since $F_C$ indicates the productive system complexity which not necessarily is already reflected in its Average Wage level.
% While, B) of Fig. \ref{us-cavw-savw-f} displays the quite reasonably positive relation between Sector Complexity and Sector Average Wage.  Given those results, we trust that our metric actually captures important and reasonable information on  the studied system.
% \textcolor{red}{ARTICOLARE MEGLIO}

\subsection{A measure for the inequality between sectors}
Analyzing the structure of the wage distribution by economic sector requires a suitable metric to determine its dispersion.  Here,  we adopt an \textit{ad hoc} Theil index  \citep{theil67}, a measure of dispersion across income distributions defined by analogy with the Shannon Entropy \citep{shannon48}.
The Theil index is defined for a population of $n$  individuals and for a discrete income distribution $\mathbf{y} \in \mathbb{R}_+^{n}$, where every $p$ individual has income $y_p$ for $p=1,...,n$, and algebraically is:
\begin{equation}\label{eq:theil}
    T=\frac{1}{n}\sum_{p=1}^n \frac{y_p}{\mu} \cdot \log\left( {\frac{y_p}{\mu}} \right) 
\end{equation}
where $\mu = \overline{\mathbf{y}}$ is the average income. $T\in[0, log(n)]$ and  is a monotonically increasing function with its upper bond dependent on the population size. When $T=0$ there is perfect equality, the situation in which everyone has the same income $\mu$. Instead, when  $T= log (n)$ inequality reaches its maximum and one individual owns all the income. Thus, by drawing this analogy with the Shannon Entropy, an inversion of extremal situations takes place. In fact, a state of maximum disorder corresponds to the minimum of the Theil index.  While there is minimum disorder when $T$ is maximum.  
%Dividing $T$  by $\log(n)$ can normalize the equation to range from 0 to 1.
% The Theil index measures an entropic "distance" the population is away from the "ideal" egalitarian state of everyone having the same income. The numerical result is in terms of negative entropy so that a higher number indicates more order that is further away from the "ideal" of maximum disorder. Formulating the index to represent negative entropy instead of entropy allows it to be a measure of inequality rather than equality.

% An important characteristic unique to these entropy based measures, immediately noted by Theil for the special case of
% his T and L indices, is that they are decomposable across groups that parse the individuals of the population into mutually exclusive, completely exhaustive, “bins”. Overall inequality can be separated into a between group component and a within group component. An important characteristic of entropy-based indexes such as the Theil index is that they are decomposable. If individuals are grouped in a mutually exclusive, completely exhaustive way, overall inequality can be separated into a between-group component and a within-group component. One of the advantages of the Theil index is that it is a weighted average of inequality within subgroups, plus inequality among those subgroups. For example, inequality within the United States is the average inequality within each state, weighted by state income, plus the inequality among states.

Entropy based measures are decomposable across population groups, this property was originally explored by Theil himself in 1967. Here we follow the approach and the formalism adopted by the \textit{University of Texas Inequality Project}. 
Indeed, if $Y=  \sum_{p=1}^n y_p$ is the total income of the population, we can rewrite Eq.\ref{eq:theil} as it follows:
\begin{equation}\label{eq:theil-sector}
T = \sum_{p=1}^n \frac{y_p}{Y}  \cdot \log \left( \frac{y_p}{Y}  \middle/ \frac{1}{n}\right).
\end{equation}
As pointed out in \citep{utip14}, expressing $T$ in the form of Eq.\ref{eq:theil-sector}: \vi  highlights a possible intuitive interpretation of the Theil index as a direct measure of the discrepancy between the distribution of income and the distribution of individuals between mutually exclusive and completely exhaustive groups''\citep{utip14}. Grouping all the individuals in $m$ groups  --each group $i$ ( $i=1, \ldots , m$) having $n_i$ individuals and total income  $Y_i$-- we can decompose overall inequality into two components  \citep{conceicao98,utip15}:
\begin{equation}
T = T'_g+T^{w}_g.
\end{equation}
$T'_g$ is the between-group component and it is given by:
\begin{equation}
T'_g = \sum_{i=1}^{m} \dfrac{Y_i}{Y} \log \left( \dfrac{Y_i}{Y} \middle/ \dfrac{n_i}{n} \right).
\end{equation}
And $T^{w}_g$, the within-group component, is given by:
\begin{equation}
T_g^{w} = \sum_{i=1}^{m} \dfrac{Y_i}{Y} T^{(i)} 
\end{equation} 
where $T^{(i)}$ is the Theil index for each group $i$ and accounts for the inequality between the members of group $i$.
Several decomposition choices are possible, for instance one might concentrate on population characteristics --such as gender, age, race, economic sector of employment and so forth-- or one might divide the population on the basis of their geographical residence . %For example inequality within United States can be seen as the sum of the inequality within-counties and between-counties plus the inequality within and between states plus ...
% Conceicao 1998: There are several reasons why it may be of interest to have a decomposable measure of inequality. One might be interested in analyzing the functional distribution of income according to some criterion that divides the overall population into groups. Examples are race, gender (both of which were explored by Theil in 1967), education level, economic sector, age, to name a few. Another reason might be associated with
% geography (different regions, like, say, states or countries, which were explored also by Theil in 1967). Another possibility is study differences in urban vs. rural populations. Yet another reason may be related to the differentiation of sources of income. 
In this paper, as in \citep{utip43}, we will look at the dispersion of wages between industrial sectors. However in our case, the geographical units are counties. In general, over a certain period of time and in a specific geographical unit, we consider a partition of the working population in ${N_{\text{s}}}$ mutually exclusive and completely exhaustive groups, with ${N_{\text{s}}}$ being the number of economic sectors present in the society under study. If ${P}$ is the total number of workers in the considered area and $\mu$ the average wage of all jobs, ${y_i}$ the average wage of the $i$-th industrial sector and ${p_i}$ the number of workers in sector $i$, then the between-sector wage inequality can be measured by a $T'_g$ of the form:

\begin{equation}
T'_{sectors} = \sum_{i=1}^{N_s} \dfrac{p_i}{P}\,\dfrac{y_i}{\mu}  \log \left( \dfrac{y_i}{\mu} \right)
\end{equation} 

% A sector’s Theil element will be positive or negative, depending on whether the sector’s average earnings are greater or less than the national average; sectors with exactly the average income make zero contribution to inequality. 
$T'_{sectors} $ decreases when average wages in low-paying sectors rise,  when they fall in high-paying sectors, or when sectors that are far from the overall average in either direction lose employment, and it increases in the opposite cases. 
$T'_{sectors} $ is a group-based measure of the dispersion in the distribution of wages. It does not describe differences between individual workers, but it measures the inequality that results from the difference in average labor income between economic sectors and it is not sensitive to within-sector wage variability. Nevertheless, thanks to  the decomposability properties of  Theil measures, it has been shown that  under some formal criteria a between-group Theil statistic also captures major characteristics of the evolution of pay inequality within industrial sectors and tracks the general movement of overall inequality in household incomes \citep{conceicao98,utip43}.

 Here, at a country-level, we employ the Theil measure developed by the \textit{University of Texas Inequality Project}, while at a county-level we compute a between-sector Theil component for U.S. counties.
% %However,  to prove the robustness of our inequality measure at a county-level, in Appendix blabla we build a Gini Index of pay inequality  across industrial sectors and the results show a similar pattern.
Notice however that, while we chose the Theil index for its decomposition characteristics, Cowell et al. show that no inequality measures can be perfectly decomposable without being highly effected by single outlying observations \citep{cowell96}. To check for the eventuality that our results are driven by few spurious observations, we will replicated our exercise using the Gini index, a measure particularly stable to single outliers \citep{cowell96}, where possible. The trends found with the Gini index are comparable to the one with the Theil index which we than consider a satisfactory measure of between-sector inequality.

\section{Results}
We intend to investigate the features of the complex relation that links the distribution of income to the process of industrialization. Industrialization causes tumultuous changes in the shape of a society and in the structure of income distribution among its population. 
As we already mentioned in Section \ref{variables}, the main novelty of our paper is in the identification of such process by considering Fitness as complementary to a more classic monetary measure, per capita GDP. We carry out a comparative study of industrialization levels among countries, in fact we employ GDP per capita of a country relative to the average GDP per capita of all the countries under consideration. Moreover, by following \citep{pugliese15}, we introduce an index useful to to study industrialization as a monodimensional process. 

%We hypothesize that country Fitness, being a measure of the complexity level and manufacturing capabilities of an economy, may be a relevant determinant of retribution dispersion. %In particular, we surmise that Fitness can give additional information on wage inequality dynamics and grasp aspects of the job market and its regulation, from technological development to worker qualification levels. 
We will take into consideration a recently introduced measure of development, the Comparative Development Index (CDI). The CDI we will use is a linear combination of Fitness ranking and relative GDP per capita logarithm for country $C$ at time $t$:
\begin{equation}\label{cdi}
CDI_{C,t}=  \beta \, Ranking( F_{C,t}) + (1-\beta) \, log(GDPpc^{rel}_{C,t}) 
\end{equation}
with a suitable $\beta$. 

Notice that, with respect to \citep{pugliese15} and another work about the use of Fitness to forecast per capita GDP levels \citep{cristelli15}, we are here using here the ranking of Fitness instead of its value. This is due to the fact that, how it is visible in figure \ref{fig:us-mcs-cavw-f-savw-q98}, the shape of matrix $M_cs$ does not allow to compute the values of Fitness but only their rankings \citep{pugliese14}.

%deve essere più chiaro

The second part of the task is the identification of inequality: different theoretical arguments relate to different empirical measures of inequality. Indeed, income has two sources, labor and capital. Labor income consists in all the labor-related earnings; while capital income is the sum of all income derived from owning capital, such as capital gains, royalties, dividends, rents, interests etcetera. 
In this paper, we will focus on two inequality dimensions: a more classical one, the capital share of total income, and another that is most common in contemporary literature, the inequality in the distribution of labor income.
In a certain time and in a certain country, the capital share of income is the ratio of capital income to total income (the sum of labor and capital), thus it constitutes a measure of the importance of capital compared to labor in a certain society. The relationship between inequality and the composition of earnings among returns on capital and returns on labor is the heart of the classical approach to inequality, from \citep{marx} to \citep{piketty-capital}, and it is seen as the outcome of a conflict between different agents: those owning capital, and those owning labor power.

Another approach in looking at the inequality of a society is to investigate the dynamics of income distribution, whether it is total income, labor income or capital income. Historically, the quantitative estimate of the statistical dispersion in the income distribution favored the construction of descriptive indexes, such as the Gini coefficient or the Theil Index. Since the fifties, with the seminal work of Simon Kuznets, this kind of inequality has been associated with the economic development and the industrialization of a country. Kuznets theorized that in the course of a country's industrialization income inequality would at first increase and, afterwards, in advanced phases of capitalist development, would mechanically decrease following an inverted \vi U'' curve \citep{kuznets55}. More recently Galbraith, studying the relation between the level of wage inequality in manufacturing and the level of income on a global scale over the years 1963--1999, finds global features in the movement of inequality, but fails to observe the entire Kuznets' curve. Nevertheless, he observes a strong negative correlation between the Theil measure of wage inequality that he developed, the UTIP-UNIDO coefficient, and GDP per capita; and he observes that there is an upswing of inequality fir the richest countries of the world in the right tail of the curve. According to Galbraith, these findings were to be expected since in the observed historical period most of the countries had already overcome the transition toward industrialization \citep{galbraith07}.
%NB DIRE DELLA AUGMENTED K CURVE DI GLABRAITH CHE TROVIAMO PURE NOI VD UTIP 17 (POCHI PAESI MOLTO RICCHI RISING INEQUALITY)

The remainder of this section is composed of two parts. In Section \ref{sec:world}, we will study the relationship between development and inequality by looking at a cross section of countries. In Subsection \ref{sec:US}, instead, we will focus on a single country, the United States,  to check if the relationships found in Subsection \ref{sec:world} between development and inequality hold at a different scale.
%, instead, we will focus on a single country, the  United States, with two goals: i) to check if the relationships found in Subsection {sec:world} between development and inequality hold at a different scale; ii) to look at the interactions at such a scale to negate differences in policy.
\subsection{A global analysis: wage inequality among countries}\label{sec:world}
We conduct a Kuznets-like analysis by considering the two aforementioned dimensions of inequality and GDP per capita, Fitness or the Comparative Development Index as industrialization measures. We will explore these relationships through a continuous non parametric description \citep{nadaraya} and, in order to examine time and space dimensions simultaneously, we pool all countries and years. The confidence bands of all the bidimensional relations are evaluated with a non parametric bootstrap-approach.

\begin{figure}[H]
\begin{center}
\includegraphics[width=70mm]{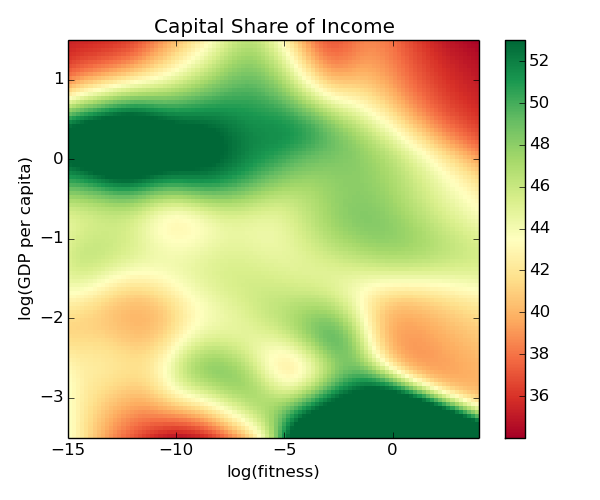}
\end{center}
\caption{The plot shows a pooling of 3059 country yearly observations over the period 1963--2000, out of a sample of 97 countries. The color map, obtained with a multivariate non-parametric kernel estimation, is a smoothed graphical representation of the average Capital Share of Income for countries with different values of Fitness and relative GDP per capita. The central green band, spanning the diagonal area from the top-left corner (countries with low fitness but high GDP per capita) to the bottom right corner (countries with high fitness but low GDP per capita), identifies the countries that are at the apex of the industrialization process. \label{fig:CapSh}}
\end{figure}

At first, we analyze the variation of the Capital Share of Income by looking at its tridimensional non-parametric kernel estimate for a country given its Fitness level and its GDP per capita relative to the other countries. 
The dependent variable, the Capital Share of Income, is graphically visualized through color variations in a color-map. As shown in Fig.\ref{fig:CapSh}, the green areas of the phase space are populated by those countries whose capital share is higher than average, up to $50\%$, while in red areas are situated countries with lower capital share, near to $30-35\%$ the average share for developed countries.

% \begin{figure}[H]
% \includegraphics[width=65mm]{UTIP-PWT-FEXP-CAPSHARE/IIvsAlpha_2D.png}
% \caption{}
% \end{figure}

\begin{figure}[H]
%\centering
\subfloat[]{
\hspace{-20pt}
  \includegraphics[width=65mm]{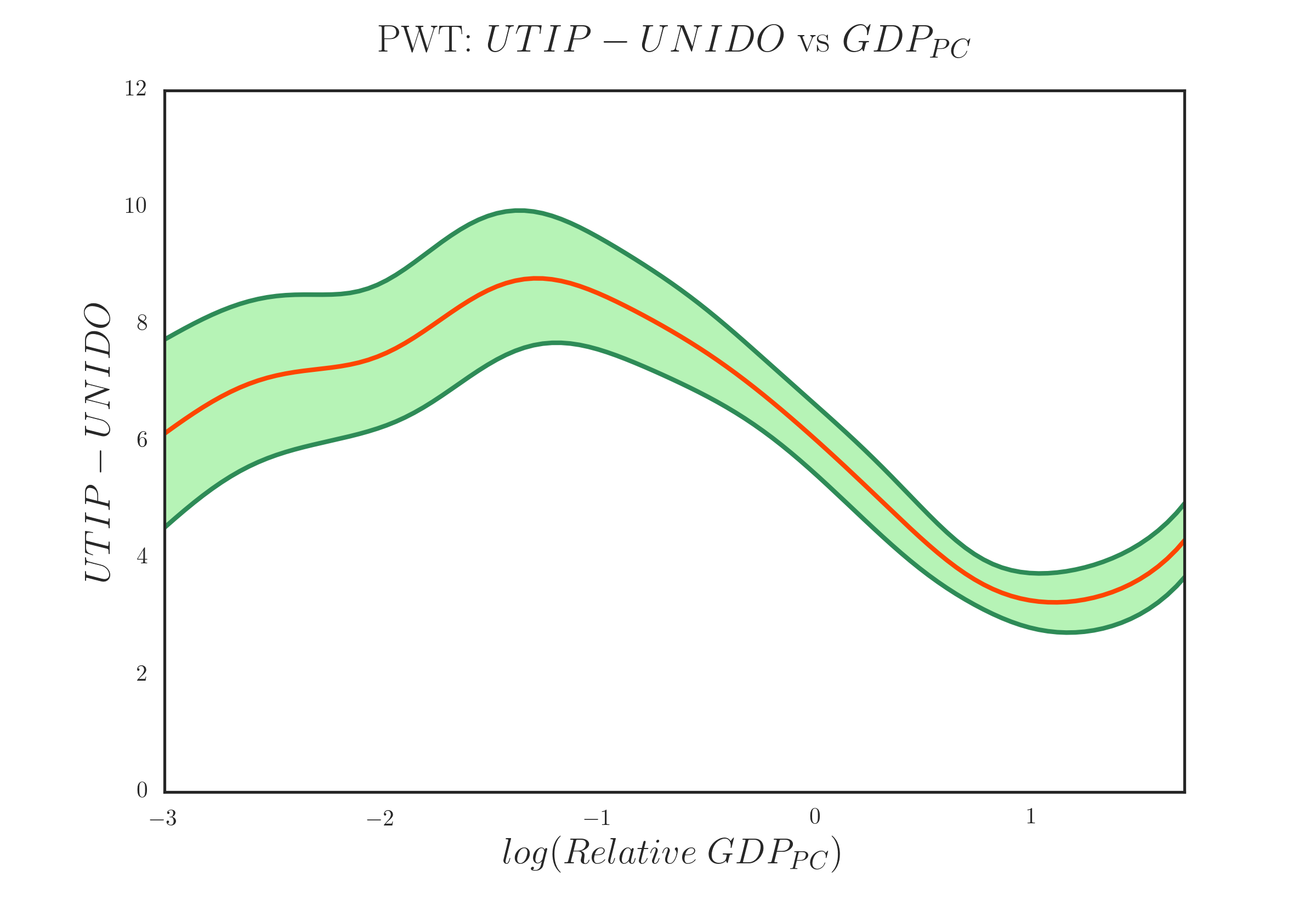}
}
\subfloat[]{
\hspace{-20pt}
  \includegraphics[width=65mm]{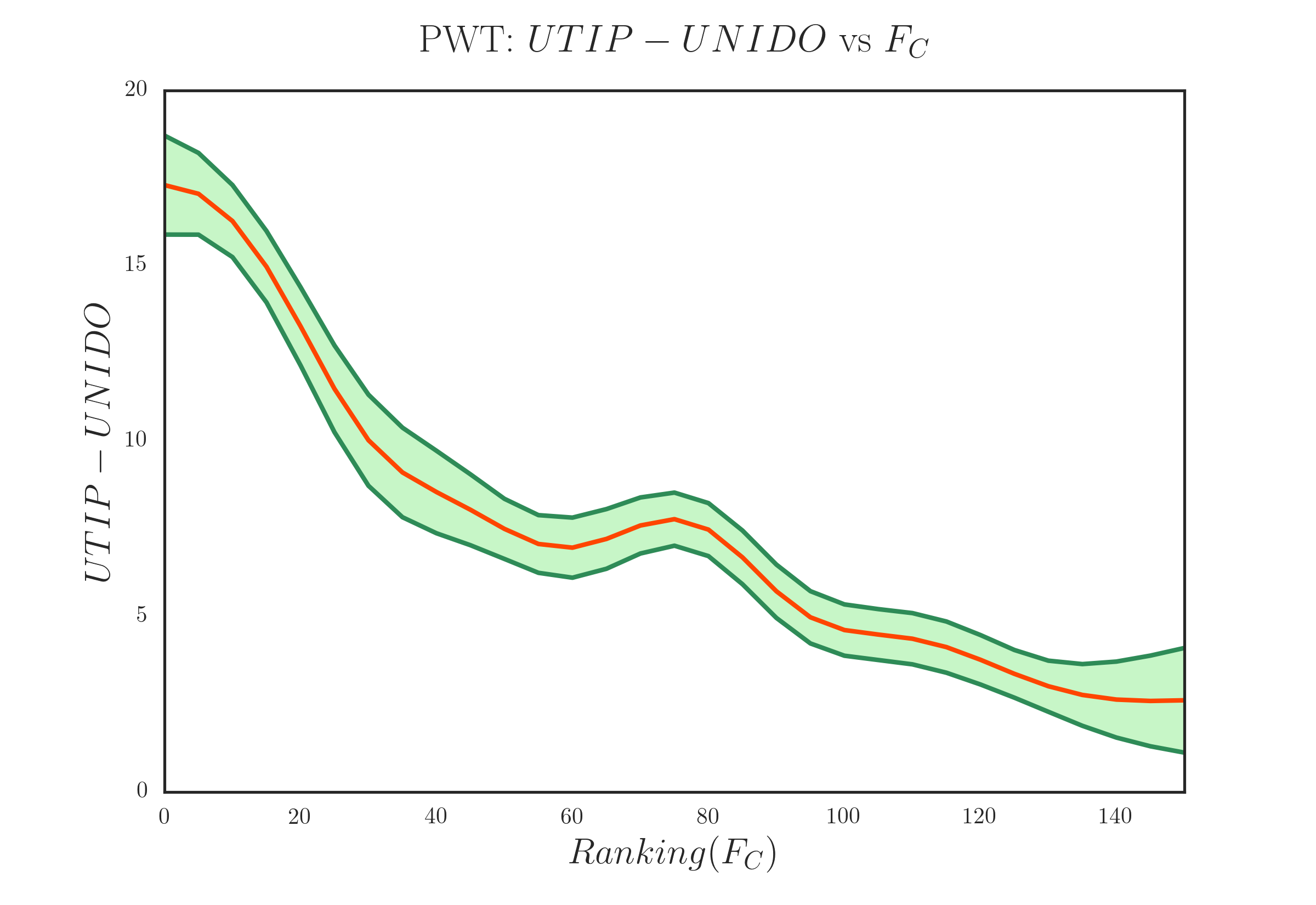}
}
\caption{\small The red lines show a non-parametric kernel estimation of the UTIP-UNIDO coefficient versus Relative GDP per capita or Export Fitness. The green shadowed areas represent a 90$\%$ confidence interval of UTIP-UNIDO expected values and have been computed with bootstrap. For the time interval 1990--2008 and for a number of countries varying between 145 and 148, we pool all countries and years for a total of 936 observations. 
\\(a): UTIP-UNIDO pay inequality measure versus relative GDP per capita. 
The negative trend reflects the one foreseen by the second half of Kuznets' curve: industrially advanced economies with high GDP per capita have low UTIP-UNIDO and vice versa. (b): UTIP-UNIDO pay inequality measure versus Export Fitness. The curve is still downward sloping and it is comparable to the one obtained in (a). } \label{utip-gdp-f}
\end{figure}

\begin{figure}%[H]
%\centering
\begin{center}

\begin{center}

\end{center}
\subfloat[]{
\hspace{-15pt}
 \includegraphics[width=65mm]{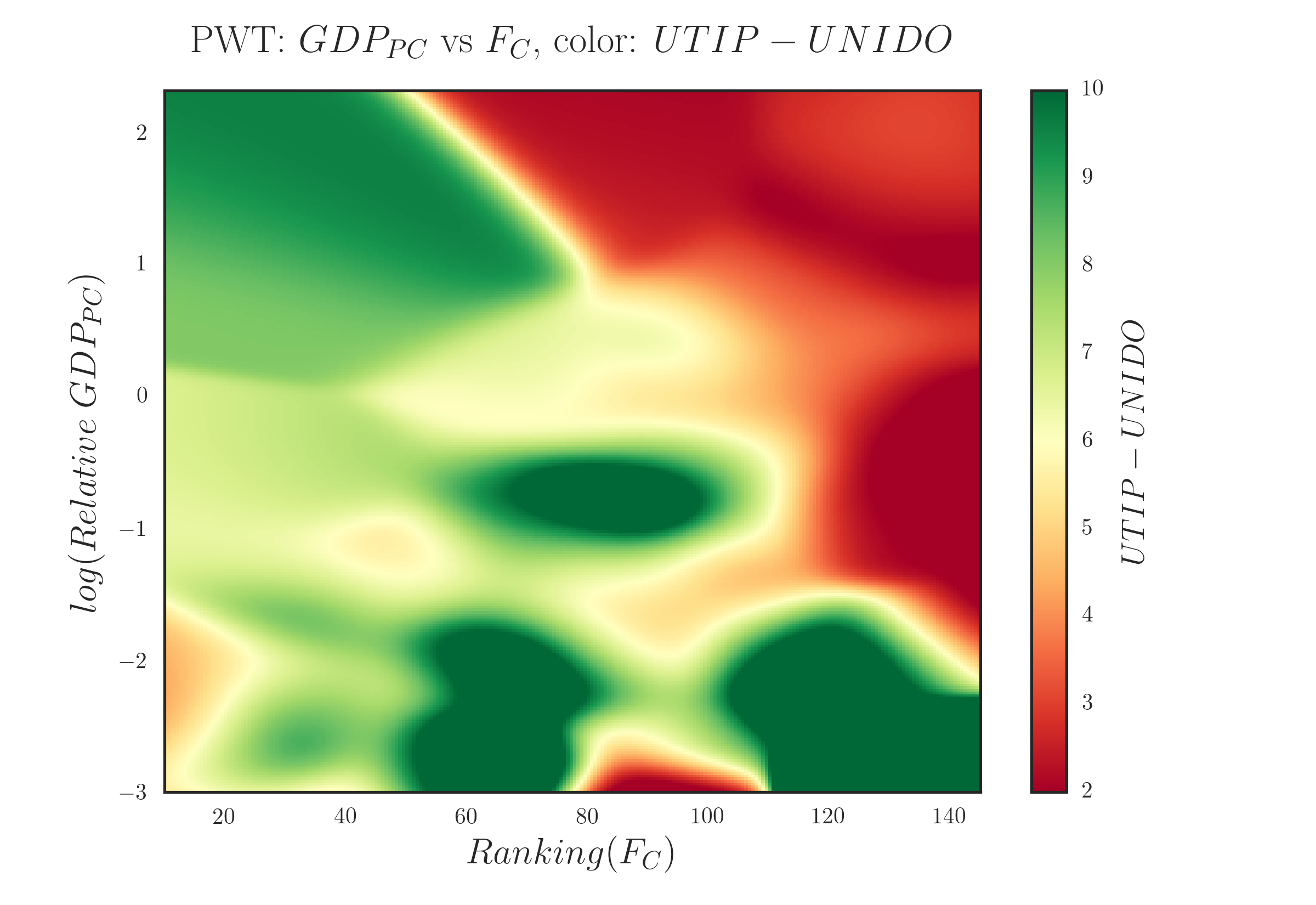}
}
\subfloat[]{
\hspace{-30pt}
  \includegraphics[width=65mm]{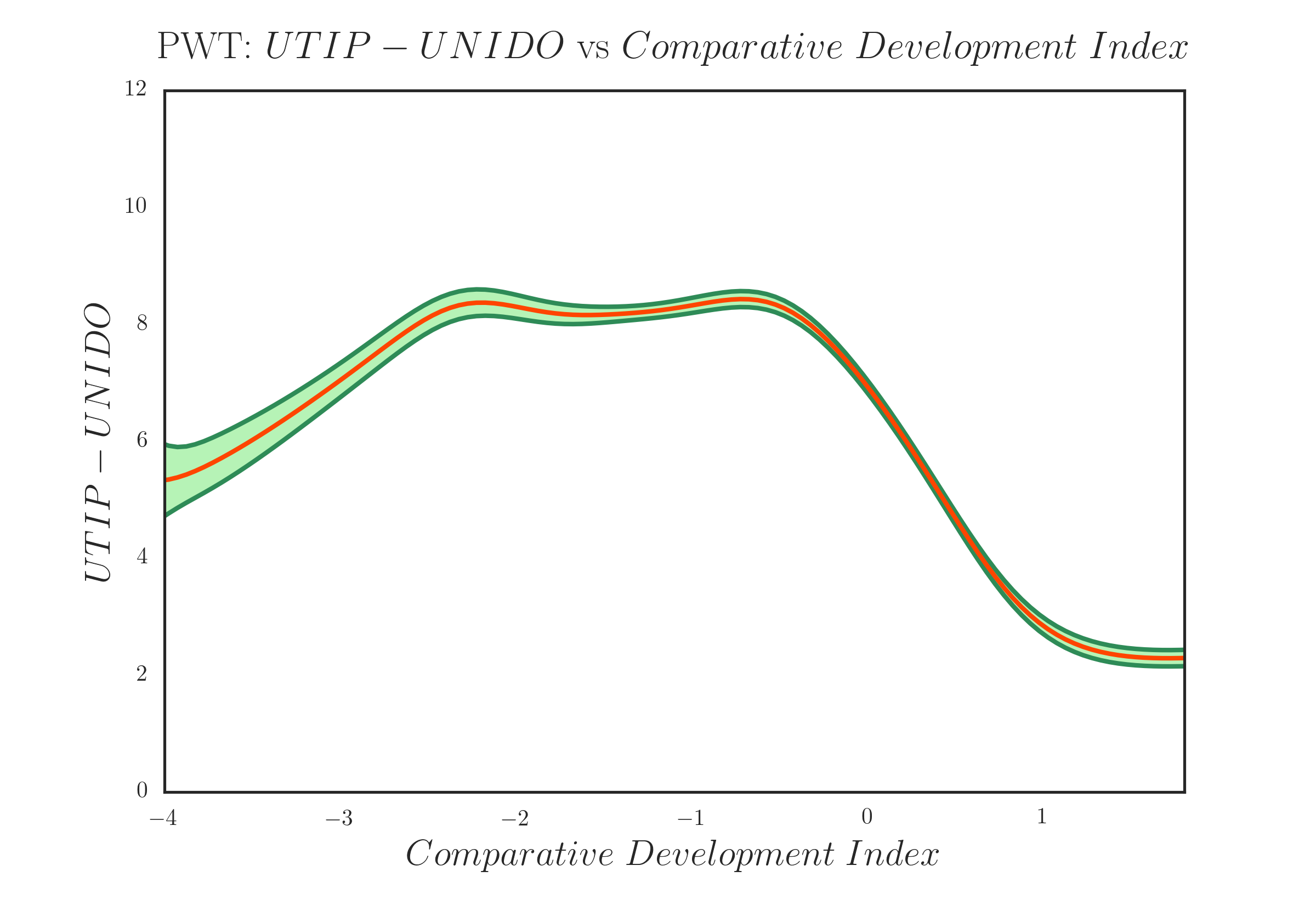}
}
\end{center}

\caption{\small Pooling of all countries and years for a total of 936 observations, over the time interval 1990--2008 and for a number of countries varying between 145 and 148. (a): A tridimensional study of UTIP-UNIDO coefficient as a function of country Fitness and Relative GDP per capita. The plot shows a pooling of all the countries and the years over the time window 1990--2008. The color-map, obtained with a multivariate non-parametric kernel estimation, is a smoothed graphical representation of the UTIP-UNIDO coefficient for different values of the country Fitness and GDP per capita. The diagonal variability of the color suggests that wage inequality between sectors, at this scale, is determined both by Average Wage and Fitness and follows a pattern similar to the one predicted by Kuznets. (b): The relationship between UTIP-UNIDO coefficient and Comparative Development Index. The red lines show a non-parametric kernel estimation and the green shadowed area represents a 90$\%$ confidence interval of UTIP-UNIDO expected values and have been computed with bootstrap. By employing the CDI as an industrialization proxy, we recover the entire Kuznets' curve.} \label{utip-gdp-f-cmap}
\end{figure}

The plot shows how the Capital Share is maximal (between 45\% and 50\%) at the moment of industrialization, and it reduces for developed countries. This is visible by looking at the industrialization as a variable depending on the combination of fitness and GDP per capita. 
Our findings do not constitute a direct reply to Piketty's claim that the Capital Share is increasing in developed countries -- a result however confirmed by our computation -- although, as a direct consequence of the low demographic and economic growth of the developed countries, they address his claim that such increase was inevitable. Indeed, while the developed countries experienced lower demographic and economic growth of the developing countries, their capital share has been lower.
% Capoverso su perché passare ad analizzare disuguaglianza redditi da lavoro.
%  The Fitness of a country, being a measure of the complexity level and manufacturing capabilities of an economy, might be able to capture the dependency of labor inequality to the level of industrial development and grasp many aspects of the job market and its regulation, from technological development to worker qualification levels. 
% EMANUELE: ORA DICIAMO CHE MEGLIO WAGE CON FITNESS, MA IN REALTA' TROVIAMO ANDAMENTI SIMILI PURE PER CAPITAL SHARE, COME GIUSTIFICARE MEGLIO?
\\ \\
From now on, we aim our attention to the movement of wage inequality. The choice of wage inequality appears more pertinent to the scope of our analysis. In fact we hypothesize that country Fitness, precisely for how it is defined, is intrinsically connected with the organization of the labor market and the sectoral division within it, and with the technological development of an industrial system. Therefore, Fitness or its combination with per capita GDP may put into light unexplored aspects of the distribution of wages.
Worldwide, to measure wage inequality we employ the UTIP-UNIDO coefficient over the period 1990--2008. Fig.\ref{utip-gdp-f} shows in section (a) the trend of wage inequality with relative GDP per capita and in section (b) with Fitness. The red lines in the figure, evaluated with non-parametric Gaussian kernel estimations, show the expected value of the UTIP-UNIDO coefficient. 
In Galbraith built the same plot as Fig.\ref{utip-gdp-f} section (a), but using data from 1963 to 2008, and found the same downward sloping relation \citep{galbraith01,galbraith07}. He argued that the trend does not follow an entire Kuznets curve but only its right side because, during the investigated time interval, much of world's countries had already experienced the first stages of industrialization.
The relationships between Export Fitness and the UTIP-UNIDO coefficient in \ref{utip-gdp-f} section (b) is still negative and is comparable to the previous one. The similarity of the dependencies might be the result of the strong correlation between Fitness itself and GDP per capita. In order to disentangle the two variables, in section (a) of Fig.\ref{utip-gdp-f-cmap}, 
%To further investigate our initial hypothesis, we need to see the effects of GDP per capita and Fitness on labor inequality simultaneously.
%an analytical tool that makes possible to disentangle F from GDP per capita and that let us identify the different features of UTIP-UNIDO dependency from GDP per capita and from Fitness.
we choose again a non-parametric continuous description and we represent the kernel estimate of the UTIP-UNIDO coefficient with a color map. Since, as comes into view in section (a) of Fig.\ref{fig:us-mcs-cavw-f-savw-q98}, the relation between Fitness ranking and relative GDP per capita logarithm is approximately linear, in section (b) of Fig.\ref{utip-gdp-f-cmap} we use the recently introduced Comparative Development Index -- as defined in Eq.\ref{cdi} -- to represent industrialization. %Nevertheless, differently from \citep{pugliese15}, we take the Fitness ranking and not its values, ...which is ok why... \citep{pugliese14}.

\begin{figure}[H]
\begin{center}
\includegraphics[width=65mm]{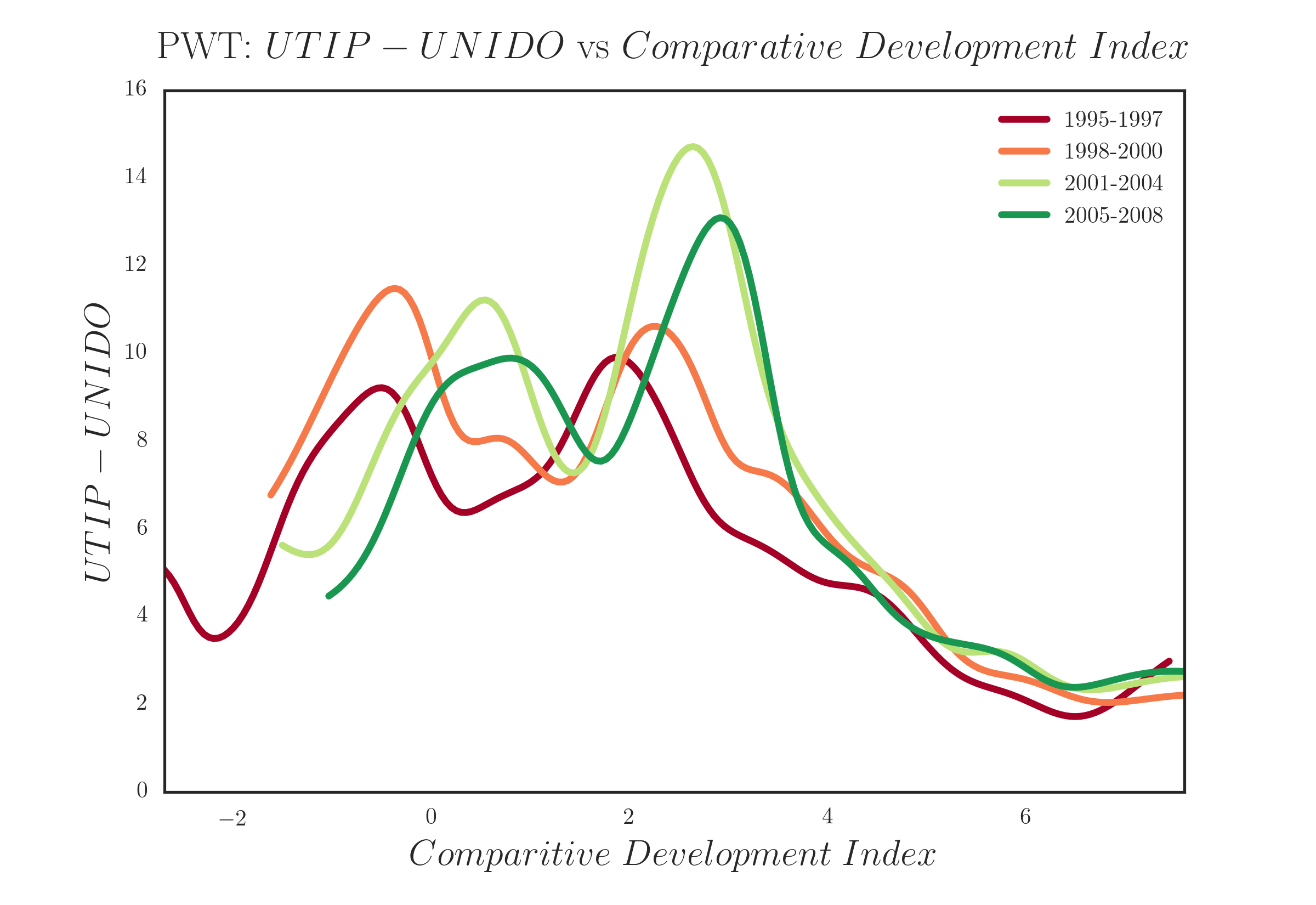}
\end{center}
\caption{UTIP-UNIDO coefficient versus Comparative Development Index. Pooling of all countries and years for the four time intervals: 1995--1997, 1998--2000, 2001--2004 and 2005--2008. The colored lines show a non-parametric kernel estimation of UTIP-UNIDO expected values. The shape shown in Fig.\ref{utip-gdp-f-cmap} section (b) is preserved over the chosen time intervals. \label{fig:CDIEvolution}}
\end{figure}

By describing industrialization as a process which concerns both the complexity and the wealth of nations, it is possible to uncover a Kuznets like pattern that was not visible in Fig.\ref{utip-gdp-f}. In fact, both in the tridimensional and in the monodimensional representation, in Fig.\ref{utip-gdp-f-cmap} it is recovered a pattern which corroborates Kuznets' hypothesis, as it is clear in section (b). %shows a pattern that seems to confirm the truthfulness of the Kuznets' curve.
Moreover, into section (a) of Fig.\ref{utip-gdp-f-cmap} in the lower-left corner are placed the industrially underdeveloped countries, with low Fitness and GDP per capita, which show intermediate and low UTIP-UNIDO values (UTIP-UNIDO $\sim  [5.5, 7.5]$). While for industrially advanced countries, which are situated in the upper-right corner, pay inequality is the lowest (UTIP-UNIDO $<4$). Ultimately, the green top-right to bottom-left diagonal is characterized by higher inequality (UTIP-UNIDO $>8.5$). In this region for industrializing countries, with low GDP per capita and high Fitness, UTIP-UNIDO takes the highest values, reaching a maximum around 49. While for countries characterized by low Fitness and high per capita GDP, as raw material exporters, UTIP-UNIDO $\sim  [5.5, 7.5]$. The role of Fitness seems indeed complementary to the role of GDP per capita in describing the level of industrialization of the country and its progress on the Kuznets curve, consistently with \citep{pugliese15}. 
%In the following we take $\beta = 0.2$.
%Fitness of a country describes the present state and the potential of its industrial systems, its complexity, it could have been more sensible to the wage's variability than the average wage. 
In conclusion, Fig.\ref{fig:CDIEvolution} shows the time evolution of UTIP-UNIDO coefficient versus Comparative Development index for four time intervals spanning from 1995 to 2008. The rightmost part of the curve clearly is downward sloping. While, for all the time intervals the leftmost part, albeit it certainly is more noisy, show a positive slope. This shape is preserved during the observed years.

%Aggiungere alle conclusioni che vale anche per share of capital
On a global scale, a Kuznets like behavior comes to light, and some global features are captured in the movement of both wage and wealth inequality, and the results appear solid during the observed time interval.
Nevertheless, the observed wage inequality dynamics are certainly also due to economical, political and cultural differences and, most of all in this case, different labor market institutions of the countries that we compare. Therefore we focus on the same dynamics but within a single country, the United States. Since the US are placed in the upper-left corner of the color-map, the expected inequality trend should be negative. 
However, at this scale, where the effects of institutional factors is deeply reduced, we are not certain that we will recover the relationship between inequality and development that we found comparing countries. It is not said that within the regions of a country the relation between political and economic inclusivity and development still holds.

 \subsection{Within one country: the case of the United States}\label{sec:US}
Here we aim to capture the same relationships between wage inequality and industrialization which we studied in the previous section. However, we will now focus on employment and wages regarding the approximately 3100 counties or counties equivalents of the United States. To do so, as already mentioned in the Section \ref{sec:data}, we employ the Quarterly Census of Employment and Wages data set for the period 1990--2014 in which industries are categorized according to the NAICS system. From this data we compute a between-sector Theil component -- as defined in Eq.\ref{eq:theil-sector} -- and a measure of Fitness for each county and each year in the interval. By implementing the Fitness-Complexity algorithm described in Eq.\ref{pilrs1} with the method outlined in Eq.\ref{mrs}, in this case for counties and NAICS industrial sectors, we obtain the County Fitness from the counties-sectors matrix $\hat M$ of dimensions ${N_{counties} \times N_{sectors}}$. Counties constitute a variegated social context, in which the industrialization level, the predominant sectors, the ethnic composition of the population can vary considerably. Per contra, in the United States a relatively uniform culture dominates and most of the economic policies are made at the state or at the federal-level.  By analyzing data regarding sub-units of a single national entity, we are able to control most of those political, economical and social factors that made the comparison between countries difficult. In fact, as showed by Moller et al., among counties the institutional effects are non significant cross-sectionally, while they can have an influence over time \citep{moller09}.
% \textcolor{red}{argomenti da trattare? us più alta disug soprattuto in wages per questo scegliamo---> vedi blau e kahn 2002 che dicono anche che nel mercato del lavoro statunistense molto derogulation e flexibility quindi meno ruolo alle istituzioini rispetto all'europa, di prete su politica a livello di stato}

%legal educational and political institutions are shared by counties
%Furthermore, another obstacle in analyzing data concerning different countries, beyond those already pointed out, is that most of the time the available data has been collected by different methods, different sources, different industrial classification and time intervals etc. However, in this case, we are able to conduct a homogeneous analysis, since all the variables of interest are calculated from the same database.
We study experimental data using the same approach for every year in the period under study. The number of sectors is constant $N_{sectors}=  89$, whereas the number of counties varies over time, spanning from $N_{counties}^{`90}=  2700$ to $N_{counties}^{`14}=  3167$. 

\begin{figure}%[H]
\centering
\subfloat[]{
 \hspace{-17pt}
  \includegraphics[width=65mm]{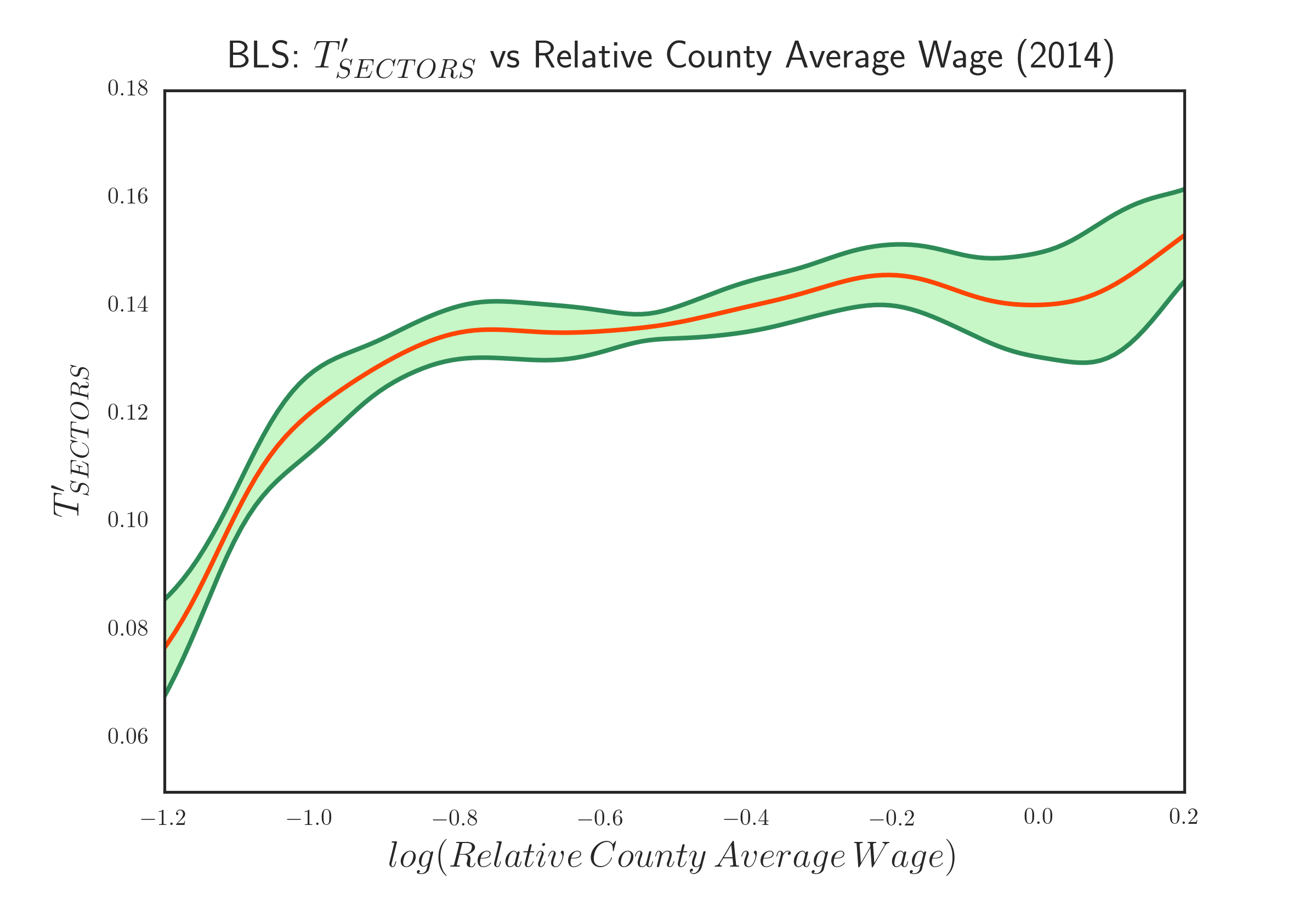}
}
\subfloat[]{
 \hspace{-17pt}
  \includegraphics[width=65mm]{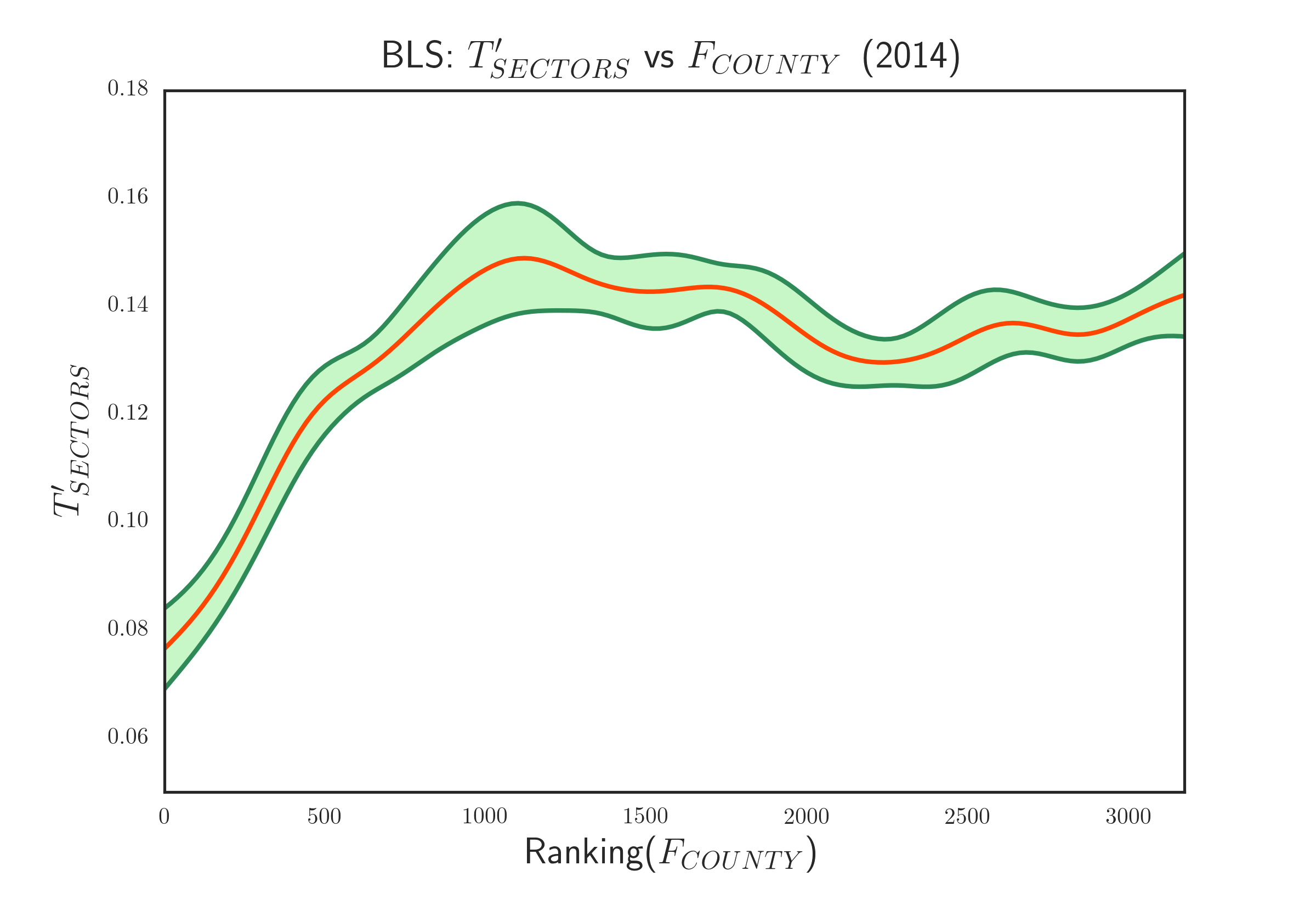}
}
\hspace{0mm}
\subfloat[]{
 \hspace{-17pt}
  \includegraphics[width=65mm]{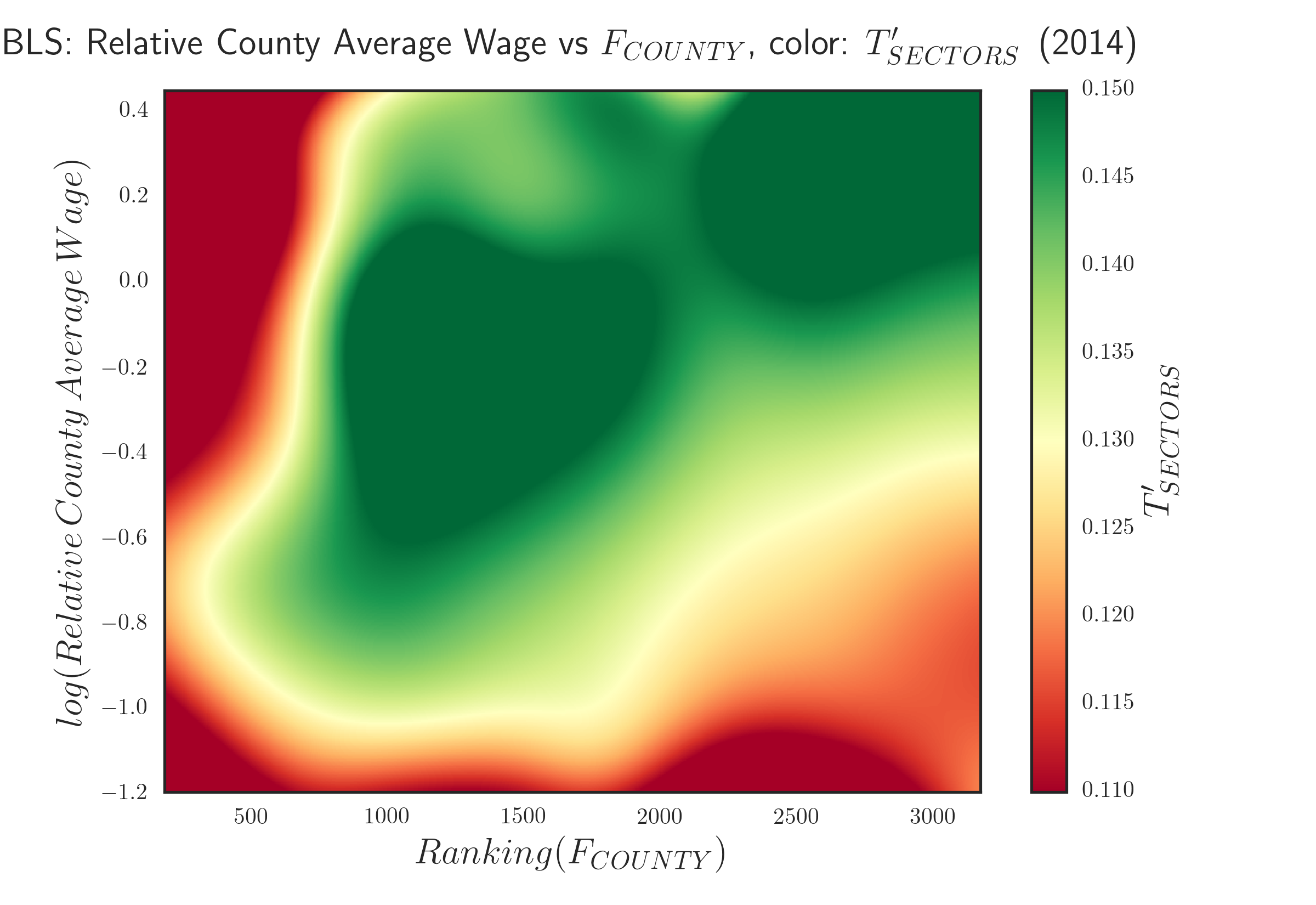}
}
\caption{\small For year 2014 and for three digit NAICS, we report the trends of the between-sector Theil component calculated from the distribution of the NAICS-sector wages.
(a) and (b): Between-sector Theil component versus Relative County Average Wage or County Fitness. We studied the relationships with non-parametric kernel regressions: the red lines depict the kernel estimation of $T'_{SECTORS}$ versus County Average Wage or $F_{COUNTY}$ and the green shadowed areas show $90\%$ confidence intervals of the expected values. The relationships are both positive-sloping, with a plateau for high $F_C$ values. (c): Tridimensional study of the county between-sector Theil index as a function of County Fitness and County Average Wage. The color-map, obtained with a multivariate non-parametric kernel estimation, is a smoothed graphical representation of the variation of $T'_{SECTORS}$. From the diagonal green band is clear that US high Fitness counties, either industrialized or in the process of ongoing industrialization, are the most unequal.
% Within a country, in this case at a county-level, the noise caused by the political and economic factors that differ between countries diminishes. In 2014, the pattern found in the cross-sectional analysis among countries is the reverse/inverse of the one between countries: as County Average Wage and County Fitness increase $T'_{SECTORS}$ soars, and so does the inequality in the wage distribution within the counties of the US. \textcolor{red}{RISCRIVERE MEGLIO}
}\label{fig:us-theil-avw-f-2014}
\end{figure}

Indeed the United States are one of the highest GDP countries, and over the two analyzed decades the American counties experienced such levels of development that, if they were to be considered independent political units, it would have been reasonable to expect them to be situated in the rightmost part of a Kuznets' curve, among developed and prosperous societies. 
%COPIATO would place many counties among the most highly developed societies if they were independent political entities
%Thus, it would be reasonable to expect to find them in the rightmost part of a Kuznets' curve.
However, as shown in Fig.\ref{fig:us-theil-avw-f-2014},  in 2014 we find an upward sloping relationship between $T'_{SECTORS}$ and $F_{COUNTY}$ or County Average Wage relative to other counties. The same pattern is visible from the diagonal green band in the tridimensional visual test which we performed in Fig.\ref{fig:us-theil-avw-f-2014} section (c). %And since, as already pointed out, County Fitness and County Average Wage are correlated, once again we perform a tridimensional visual test in order to bring to light the effect of industrialization on wage inequality variability. We put $F_{COUNTY}$ on the abscissa, County Average Wage on the ordinate and we represent $T'_{SECTORS}$ with color. Still, within the US in 2014, as we can see from the diagonal green band in Fig. \ref{fig:us-theil-avw-f-2014} section (c), as industrial development increases so does wage inequality.
%us-theil-avw-f-2014
In 2014, a county with a high Fitness or Average Wage has more likely high inequality in the distribution of wages among sectors. What we saw on a country-level was well known in the literature: in high GDP per capita countries labor income is more equally distributed (see for example \citealp{utip17}). %\textcolor{red}{FIND OTHER REFERENCES or rewrite or explain better}.
Still, looking at the same dynamics with a magnifying glass at a scale where we do not observe the effect of redistributive policies, the more complex and diversified is a productive system, the higher is the wage inequality among sectors in it. %/as industrial development increases so does wage inequality.
Thus, at net of the social and institutional factors that differ among countries, it is possible to observe that there is an upturn in the relation between wage inequality and development.% is overturned/reversed. 

%us-theil-avw-f-1990
\begin{figure}[H]
\centering
\subfloat[]{
 \hspace{-17pt}
  \includegraphics[width=65mm]{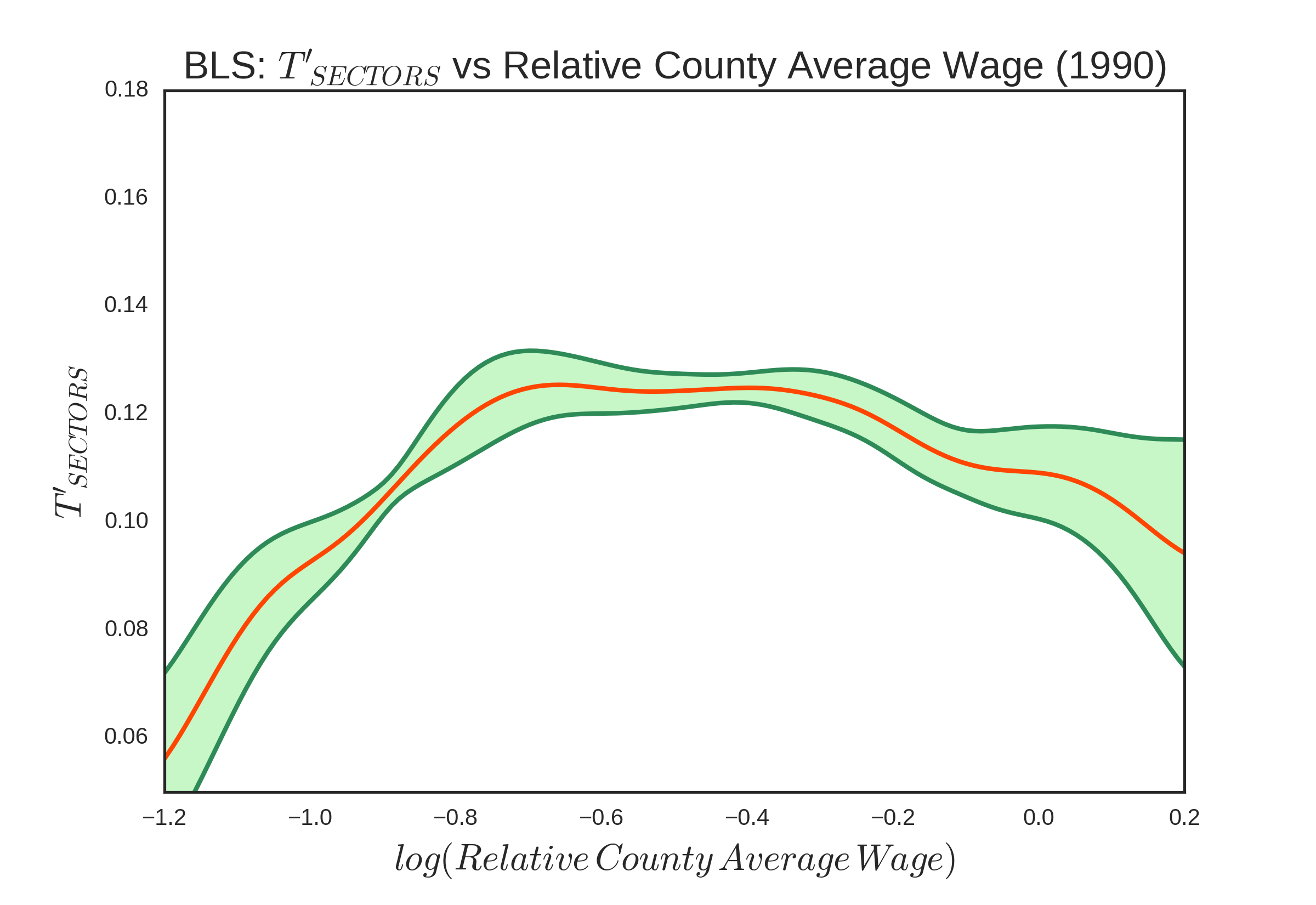}
}
\subfloat[]{
 \hspace{-17pt}
  \includegraphics[width=65mm]{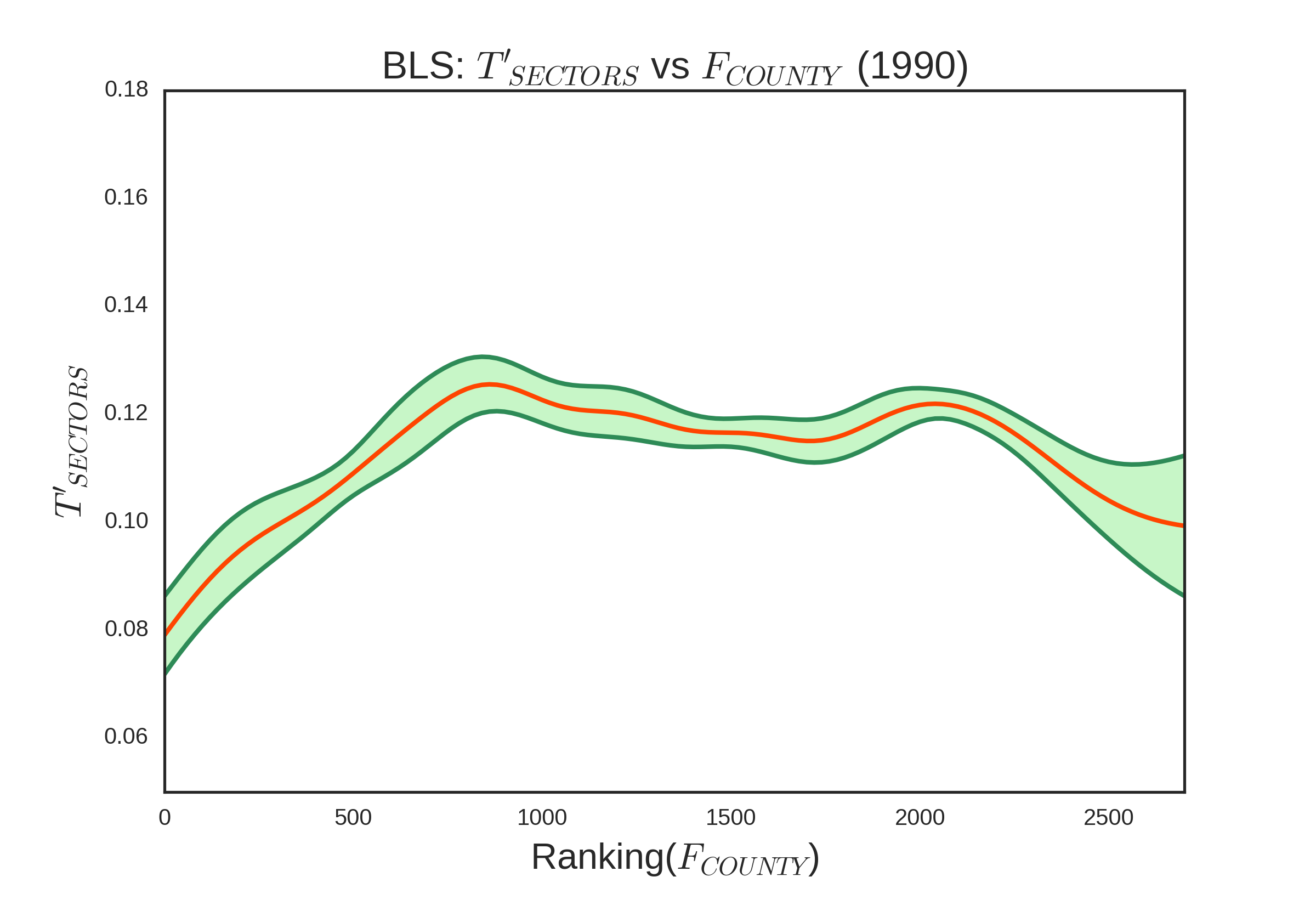}
}
\hspace{0mm}
\subfloat[]{
 \hspace{-17pt}
  \includegraphics[width=65mm]{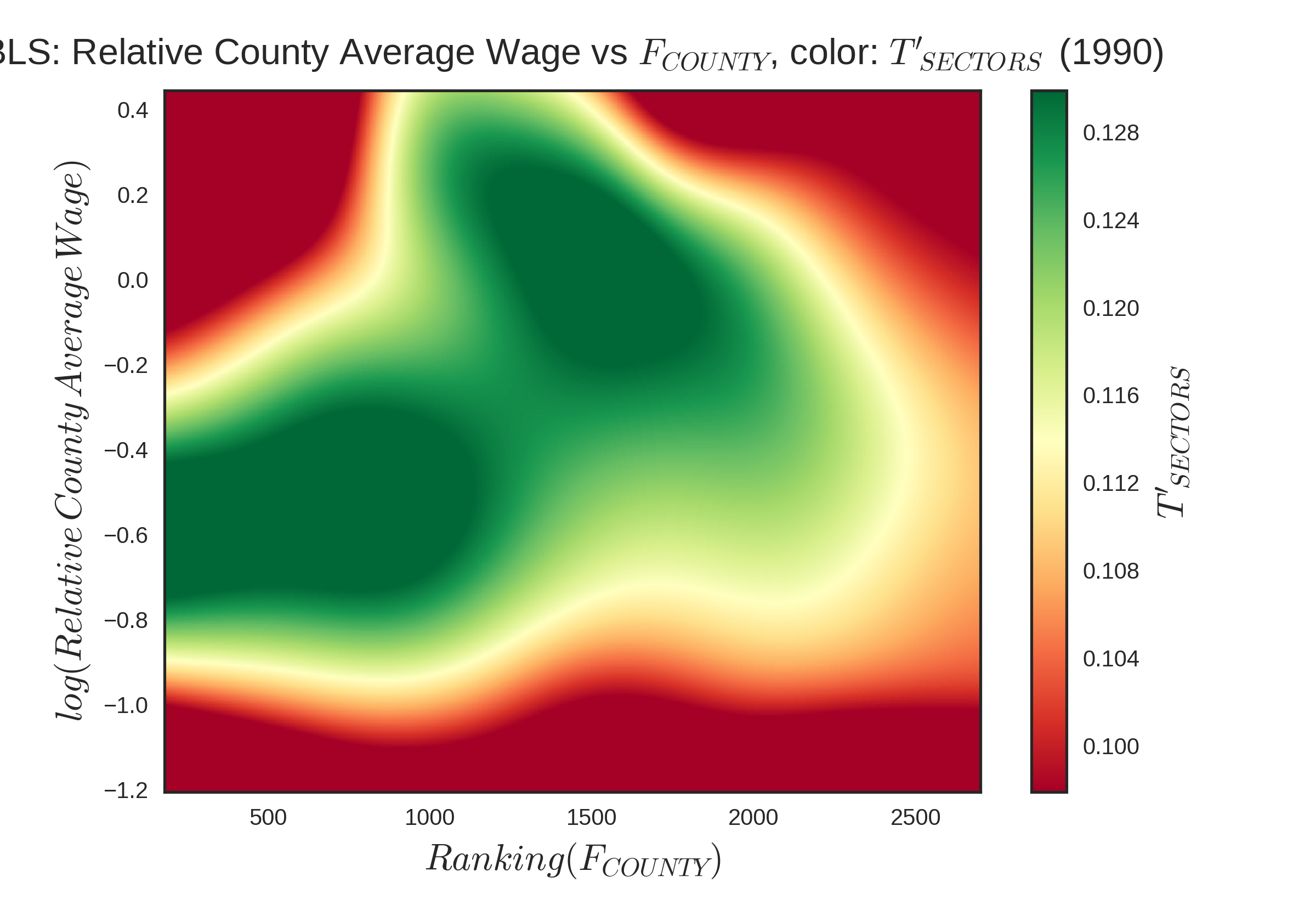}
}
\caption{\small Trends of the between-sector Theil component calculated from the distribution of the NAICS-sector wages for 1990 and three digit NAICS. All the relationships are analyzed with the same methods of Fig.\ref{fig:us-theil-avw-f-2014}.
(a) and (b): Between-sector Theil component versus Relative County Average Wage or County Fitness. (c): Color map of the variation of Between-sector Theil component as a function of County Fitness and County Average Wage. Differently from 2014, in 1990 counties' inequality follows a Kuznets-like process: it grows until a certain level of $F_{COUNTY}$ or Relative County Average Wage and then, after a small plateau, it decreases.}\label{fig:us-theil-avw-f-1990}
\end{figure}

This pattern appears from 1993 and remains stable up to 2014. In the first years of the nineties, as shown in Fig.\ref{fig:us-theil-avw-f-1990} for 1990, the curve differs from the one found in the following years. In 1990 inequality at first increases and then decreases, tracing a Kuznets' curve, a similar pattern to the one found in the global pooled analysis. 
%The plots for different years are presented in the Supporting Materials
%POTREMMO DIRE QUI CHE SE AVESSIMO DATI PRIMA DEL 1990 POTREMMO RENDERE PIÙ FORTI LE NOSTRE AFFERMAZIONI. POTREMMO CITARE L'ALTRO ARTICOLO DI NIELSEN SULLE CONTEE, QUELLO IN CUI FACEVANO VEDERE LA CURVA DISCENDENTE DELLA INCOME INEQUALITY PER 1970 E 1980 E USARLO PER RINFORZARE NOSTRA TESI (PONENDO L'ACCENTO SUL FATTO CHE NOI OSSERVIAMO ANCHE LA PARTE SINISTRA DELLA CURVA DI K. E NON SOLO LA DX CON WAGE INEQUALITY, MA CMQ È RUMOROSA LA NOSTRA PARTE CRESCENTE).

Nielsen and Alderson analyzed the movement of household income distribution  within US counties with the county average income in 1970, 1980 and 1990 \citep{nielsen97}: the three curve are all downward sloping but in 1990 there is a small inequality increase for the more developed counties which they ascribe to the great U-turn of the US inequality cycle in the last four decades. Unfortunately, since we analyze a different time interval, we cannot compare directly our results whit theirs; also the object of analysis is different, nevertheless what we saw for 1990 might be related to their 1990's observation of a turning point in the trends of household income inequality.

\begin{figure}
\centering
\subfloat[]{\includegraphics[width=65mm]{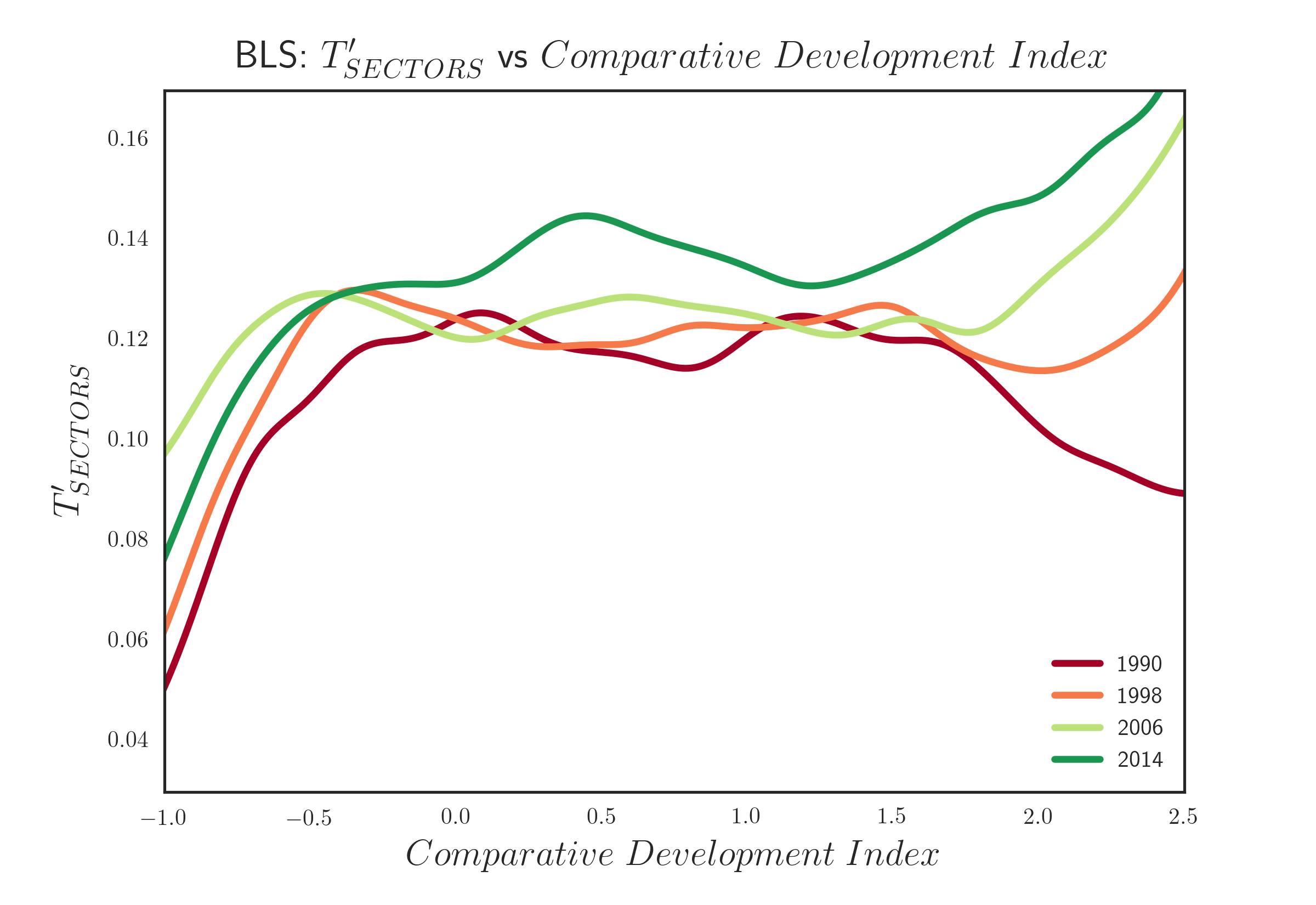}
}
\caption{\small $T'_{SECTORS}$ as a function of Comparative Development Index for the US counties in the years 1990, 1998, 2006 and 2014. We observe two main feature: (i) wage inequality grows over the time; (ii) in 1990 we found a Kuznets-like pattern, while in the following years the second half of the curve experiences a turnaround, and as County Average Wage and County Fitness increase the inequality in the wage distribution within the American counties soars.}\label{fig:us-cdi-evolution}
\end{figure}

The trend change can be distinctly observed by looking at the time evolution of the Theil index as a function of the Comparative Development Index, as in Fig.\ref{fig:us-cdi-evolution}. From 1990 (dark red curve), in which the most developed counties show declining inequality, to 2014 (dark green curve) is uncovered a reversal of matters of great significance: the trend goes from a parabola to a monotonically increasing function. So, we are not merely reporting the effect of the longitudinally upswing in wage inequality in the US, a matter explored widely and in multifaceted ways in the literature -- to name a few: \citealp{bluestone88,goldin91,katz-goldin09,autor08,piketty01us,utip57} etcetera -- but also a structural change in the distribution of sectoral wages in the last decades. In fact, after 1992 we saw a change in the relationship between development and wage inequality and, if the turnaround was merely due to politics, we would have observed it over the whole time interval. The Kuznets' curve is not the natural evolution of inequality, is just one path that can be followed by a society under certain circumstances. 
%  %development does not necessarily induce a Kuznets curve,

% %EMANUELE: 1) AIUTO SPIEGARE MEGLIOOOO ULTIMO CONCETTO
%           2) SULLA LETTERATURA SU INCOME/WEALTH/WAGE INEQUALITY IN US...
%           CI SAREBBE MOLTO DA DIRE, SIA SULLE DIVERSE ANALISI CHE SULLE 
%           CAUSE. ANCHE TRALASCIANDO LE CAUSE, SOLO SUL FENOMENO BISOGNA
%           FARE CAPPELLO A INIZIO DELLA SEZIONE US O METTERE PEZZO NELL'
%           INTRODUZIONE? COSÌ NOMI A CASO FORSE NON HA SENSO! QUELLI CHE
%           MESSO GUARDANO TUTTI WAGE/PAY

\section{Conclusion}
%This paper offers gives new information  a novel way to identify industrialization
In this paper we have examined the dynamic of wage inequality with the industrialization process stylized with a measure of economic complexity. Firstly, we focused on a global scale uncovering a Kuznets like pattern in a pooled relationship between a measure of pay inequality, the UTIP-UNIDO coefficient, and the Comparative Development Index over the time interval 1963--2008. This trend, that appears persistent longitudinally, was not visible in relation with a more classic measure of development such as relative GDP per capita. Similar results are obtained with another dimension of inequality, the Capital Share of Income.
Moreover, we have presented a generalization of the Fitness-Complexity algorithm: in the evaluation of the complexity of an economy we have considered industrial sectors and generic geographic areas by weighting the importance of a certain sector in a region by its wage volume distribution compared to a reference distribution. 
We have applied this method to US counties during the time window 1990--2014 by employing NAICS industries data on wages and employment at a th. The redefinition of the algorithm applied to the US performs well and appears to be a valid tool in the analysis of the comparative development of the US counties. We have then computed a Theil measure for inequality in the distribution of wages within counties between NAICS sectors. The US counties have proved to be an adequately heterogeneous landscape which represents quite faithfully the industrialization spectrum of developed countries.%---great geographical variation 
The Kuznets pattern that appeared worldwide is not recovered within the US, there is in fact a substantial trend reversal: at a county level wage, from the early nineties,  inequality increases with growing industrialization. 
Furthermore, the functional form of the relation between wage inequality and industrialization proxies varies over time: in 1990 it follows a Kuznets curve while in the following years a monotonically increasing trend emerges.
Within counties the effect of institutional factors has been proved to be minimal, nevertheless the trend change underlines an ongoing structural change in the movement of wage inequality. %...
To summarize: the structure of wage inequality is a non scale invariant, a single national entity behaves differently from its components. And additionally, the process within a country varies over time.
We have presented some empirical evidence on the movement of wage inequality within US counties, not with the aim of providing a satisfactory description of the determinants of such inequality but to cast new light on the long-standing debate that surrounds the relation between development and inequality. %A natural continuance of this research would be..... NON SE HA SENSO METTERE UNA FRASE DEL GENERE

Moreover, we are developing further investigations in a paper soon to be submitted on the working principles of the generalization of the Fitness-Complexity algorithm at different geographical scales and industrial aggregation levels.

\bibliographystyle{plainnat}
\bibliography{biblio.bib}

\end{document}